\documentclass[11pt]{article}

\usepackage{graphics,graphicx}
\usepackage{amsmath,amsthm,amssymb}
\usepackage[english]{babel}
\usepackage{textcomp}
\usepackage{rotating}
\usepackage{lscape}
\usepackage{longtable}
\usepackage{color}
\usepackage[latin1]{inputenc}
\usepackage{fancyhdr}
\usepackage{pst-all}
\usepackage[bookmarks = true, colorlinks=true, linkcolor = blue, citecolor = blue, menucolor = black, urlcolor = black, plainpages=false]{hyperref}
\usepackage[T1]{fontenc}
\usepackage{rawfonts}
\usepackage{amsfonts}
\usepackage{enumerate}
\usepackage{pstricks}
\usepackage{layout}
\usepackage{colortbl}
\usepackage{multimedia}
\usepackage{times}
\usepackage{array}
\usepackage{supertabular}
\usepackage{xspace}
\usepackage{ragged2e}
\usepackage{multicol}
\usepackage[active]{srcltx}
\usepackage{makeidx}
\usepackage{subcaption}
\usepackage{dsfont}
\usepackage{epstopdf}
\usepackage{breakcites}

\newtheorem{Theorem}{Theorem}[section]
\newtheorem{Definition}[Theorem]{Definition}

\newtheorem{Proposition}[Theorem]{Proposition}
\newtheorem{Property}[Theorem]{Property}
\newtheorem{Lemma}[Theorem]{Lemma}
\newtheorem{Corollary}[Theorem]{Corollary}

\newenvironment{Proof}[1][Proof]{{\text{#1.\;\;}}}{\hfill $\square$}%
\numberwithin{equation}{section} 
\numberwithin{Theorem}{section}

\begin{document}




\title{\Large{\textbf{A Directional Multivariate Value at Risk}}}\vspace{1cm}

\author{Raúl Torres (ratorres@est-econ.uc3m.es) \\
Rosa E. Lillo (lillo@est-econ.uc3m.es) \\
Henry Laniado (hlaniado@est-econ.uc3m.es)}
\date{}

\maketitle

\begin{abstract}

In economics, insurance and finance,  value at risk (VaR) is a widely used measure of the risk of loss on a specific portfolio of financial assets. For a given portfolio, time horizon, and probability $\alpha$, the $100\alpha\%$  VaR is defined as a threshold loss value, such that the probability that the loss on the portfolio over the given time horizon exceeds this value is $\alpha$.
That is to say, it is a quantile of the distribution of the losses, which has both good analytic properties and easy interpretation as a risk measure. However, its extension to
the multivariate framework is not unique because a unique definition of  multivariate quantile does not exist.
In the current literature, the multivariate quantiles are related to a specific partial order considered in $\mathbb{R}^{n}$, or to a property of the univariate quantile that is desirable to be extended to $\mathbb{R}^{n}$.
In this work, we introduce a multivariate value at risk as a vector-valued directional  risk measure, based on a directional multivariate quantile, which has recently been introduced in the literature.
The directional approach allows the manager to consider external information or risk preferences in her/his analysis.
We have derived some properties of the risk measure and we have compared the univariate \textit{VaR} over the marginals with the components of the directional multivariate \textit{VaR}.
 We have also analyzed the relationship between some families of copulas, for which it is possible to obtain closed forms of the multivariate  \textit{VaR} that we propose.
Finally, comparisons with other alternative multivariate \textit{VaR} given in the literature, are provided in terms of robustness.
\end{abstract}

%


\section{Introduction}\label{sec:intro}
Value at risk (\textit{VaR}) has become a  benchmark for risk management which is defined as the threshold quantity that does not exceed a certain probability level which is considered to be dangerous. It is commonly implemented by  investment banks to measure the market risk of their asset portfolios.
  Although  (\textit{VaR}) has been broadly criticized from the work of  \cite{artzner} since it does not verify the diversification property,  it has also been defended by \cite{heyde} for its robustness. For univariate risks, the \textit{VaR} is simply the $\alpha-$quantile of the loss distribution function. Thus, the \textit{VaR} is a risk measure easily interpretable, and it still remains the most popular measure used by risk managers. Unfortunately, a unique definition of multivariate \textit{VaR} is more complicated because there are different possible definitions of multidimensional quantiles that try to generalize some desirable properties of the univariate quantile. For instance, the proposals given by \cite{kol} of multivariate quantiles as inversions of mappings, multivariate quantiles in terms based on norm minimization as in \cite{Chaudhuri}, multivariate quantiles as level-sets given by \cite{fpsll}, multivariate quantiles based on depth functions developed in \cite{serfling2}, and finally, multivariate quantiles based on projections as in \cite{pateiro}, \cite{hapasi}, \cite{mizera}.

Currently business and financial activities generate data for which it has been shown that it is insufficient to consider single real-value  measures over marginal aspects, in order to quantify risks jointly associated to the data. For instance, one of the drawbacks detected in the global banking regulatory \textit{Basel II} is the solvency and liabilities dependence among the financial institution branches, or even the domino effect in the markets that could be generated by dependence among filial products. Thus, the solvability of each individual branch may strongly be affected, not only by its activities, but also by the level of dependence among all the branches. In consequence, it is necessary to quantify the risk, considering both the  multivariate nature of the data  and the dependence among the marginal risks.

In \textit{Basel III}, a new liquidity regulation was proposed  in order to avoid the weakness detected in  the 2007-2009 crisis; but these regulations have to be complemented by internal  models in the institutions, in order to obtain better hedge results. These models have to include multivariate risk measures computable in high dimensions and also, to consider possible internal and external risks, even if the nature of those risks is strongly heterogeneous.

In recent decades, literature devoted to extend the \textit{VaR} measure to the multivariate setting has been published. For instance, bivariate versions have been studied in \cite{arbia}, \cite{tibi2}, \cite{nap06}. Also, for multivariate distributions in general, some notions of \textit{VaR} have been introduced (e.g. \cite{prekopa,ep,bernardino}).\cite{ep} linked the risk measure to the level surface defined when the distribution function of risk $\mathbf{X}$ or the survival function accumulate some $\alpha$-value, which is considered as a quantile surface. Recently, \cite{bernardino} introduced a new notion of multivariate \textit{VaR} based on those level surfaces studied in \cite{ep}. They commented  that considering the whole surface as a risk measure could induce interpretation problems. Therefore, they defined the multivariate \textit{VaR} as the mean of the points belonging to the surface considered in \cite{ep} and hence, the output  is a point with the same dimension as the random vector of losses. Specifically, they define the {\it upper--orthant Value--at--Risk} ({\it lower--orthant Value--at--Risk}) at  $\alpha$--level ($(1-\alpha)$--level) as the conditional expectation of $\mathbf{X}$, given that $\mathbf{X}$ stands in the $\alpha$-set of its distribution (survival) function.

In this paper, we introduce a \textit{directional multivariate Value at Risk}, based on the extremality level sets introduced in \cite{laniado-lillo-romo}, which permit the concept of directional multivariate quantile to be defined. The extremality level sets are  surfaces defined by following the same idea as in \cite{ep} but linked to rotations of the  multivariate distribution; that is, a directional approach is considered.
We share with \cite{bernardino} the idea that a multivariate \textit{VaR} seen as a surface could bring problems in relation to its interpretation. Hence, we highlight  the idea of considering the multivariate \textit{VaR} as a vector-valued point  that defines the vertex of an oriented orthant in the direction of analysis. The vertex is obtained using the mean of  $\mathbf{X}$ to fix a reference system.
The risk measure that we propose considers the high dimension nature of the real problems, and the dependence among the risks is implied in the analysis. Finally, we give the possibility of considering manager preferences, introducing a parameter of direction $\mathbf{u}$. For instance, directions like the maximum variability given for the principal components in the portfolio, or the assets weight composition could be more interesting to analyze than the classic directions given for the  information summarized in the survival or cumulative distribution functions. Besides, the directional approach allows us to give bounds for the \textit{VaR} related to linear combination of random variables, mainly when they are statistically dependent. 

We have proved properties of the directional \textit{VaR} that we consider as relevant for a  multivariate risk measure, such as consistency with respect to a particular stochastic order and tail subadditivity in the mean loss direction, as well as some invariance properties. We have compared the components of the directional multivariate \textit{VaR} with the univariate \textit{VaR} on the marginals, in order to show that the vector given by the \textit{VaR} on the marginals provides incomplete information about the joint risk.

We have also obtained closed expressions of the \textit{VaR} when bivariate copulas are considered or when a multivariate Archimedean's copulas governed the dependence among the components of the portfolio.
Finally, we will present comparisons in terms of robustness with the alternative vector-valued multivariate \textit{VaR}, introduced by \cite{bernardino}.

The paper is structured as follows. In Section 2, we introduce some preliminary concepts and notation necessary in order to understand the main contributions of the paper. In Section 3, the \textit{directional multivariate Value at Risk} ($VaR_{\alpha}^{\mathbf{u}}(\mathbf{X}))$ is introduced  and  we provide analytic   properties, which can be viewed as extensions of those given in  \cite{artzner}, to the multivariate setting. Section 4 contains the comparisons between the  univariate \textit{VaR}  over the marginals and  the components of the directional multivariate \textit{VaR}. Section 5 is devoted to  theoretical results and closed forms of the multivariate \textit{VaR} when particular families  of copulas are considered.  In Section 6, we develop the robustness analysis. Finally, some  conclusions are outlined as well as some possible directions for future work.

\section{Preliminaries}\label{sec:prel}

The main objective of this paper is to introduce a directional multivariate Value at Risk, based on the notion of directional multivariate quantile given in \cite{laniado-pend}. In order to make the paper self contained, we have devoted this section to revise the main concepts that are necessary to properly define the risk measure introduced in this paper.

\begin{Definition}\label{orthant}
An oriented orthant in $\mathbb{R}^{n}$ with vertex $\mathbf{x}$ in the direction $\mathbf{u}$ is defined as,

\begin{equation}\label{convexcone}
	\mathfrak{C}_{\mathbf{x}}^{\mathbf{u}}=\{\mathbf{z}\in \mathbb{R}^{n}:R_{u}(\mathbf{z}-\mathbf{x})\geq 0\},
\end{equation}

where $\mathbf{u}\in\bar{\mathbb{B}}_{n}(0)=\{\mathbf{v}\in\mathbb{R}^{n}:||\mathbf{v}||=1\}$ and $R_{\mathbf{u}}$ is the orthogonal matrix such that $R_{\mathbf{u}}\mathbf{u}=\mathbf{e}$, with $\mathbf{e}=\frac{\sqrt{n}}{n}[1,...,1]'$.
\end{Definition}

Based on the oriented orthant concept, we can define a partial data order (denoted by $\preceq_{\mathbf{u}}$) in $\mathbb{R}^{n}$ as,

\begin{equation}\label{porder}
	\mathbf{x} \preceq_{\mathbf{u}} \mathbf{y},\quad\text{ if and only if, }\quad \mathfrak{C}_{\mathbf{x}}^{\mathbf{u}}\supseteq\mathfrak{C}_{\mathbf{y}}^{\mathbf{u}},
\end{equation}

where $\mathbf{x},\mathbf{y}\in \mathbb{R}^{n}$. Or equivalently,

\[\mathbf{x} \preceq_{\mathbf{u}} \mathbf{y},\quad\text{ if and only if,}\quad R_{\mathbf{u}}\mathbf{x}\leq R_{\mathbf{u}}\mathbf{y},\]
where the order on the right side is component-wise.

Throughout the paper we will use the following notation related to subsets in $\mathbb{R}^{n}$. Given $\mathbf{b}\in \mathbb{R}^{n}$, $c\in\mathbb{R}$, and $A\subset\mathbb{R}^{n}$, the sets $\mathbf{b}+A$ and $cA$ are defined as,

\begin{equation}\label{operationOrt}
\mathbf{b} + A:=\{\mathbf{b}+\mathbf{a}:\mathbf{a}\in A\},\qquad cA :=\{c\mathbf{a}:\mathbf{a}\in A\}.
\end{equation}

We recall some results on oriented orthants that will be useful in the main sections of the paper.

\begin{Lemma}\label{prop:oddOrt} Given a direction $\mathbf{u}$ and a vertex $\mathbf{x}$, then

\begin{equation}\label{eq:oddOrt}
\mathfrak{C}_{\mathbf{x}}^{\mathbf{u}}=-\mathfrak{C}_{-\mathbf{x}}^{-\mathbf{u}}.
\end{equation}
\end{Lemma}

The proof is given in the Appendix.

\begin{Lemma}\label{prop:invOrt} Given $c>0$ and $\mathbf{b}\in \mathbb{R}^{n}$, then

\begin{equation}\label{eq:invOrt}
	\mathfrak{C}_{c\mathbf{x}+\mathbf{b}}^{\mathbf{u}}=c\mathfrak{C}_{\mathbf{x}}^{\mathbf{u}}+\mathbf{b}.
\end{equation}
\end{Lemma}
\begin{Proof}
The proof is straightforward using the definitions given in (\ref{operationOrt}).
\end{Proof}

We also recall some definitions of useful stochastic orders; see \cite{shaked}, for more details.

\begin{Definition}\label{classicOrders}
Given two random vectors $\mathbf{X}$ and $\mathbf{Y}$, $\mathbf{X}$ is said to be
smaller than $\mathbf{Y}$ in:
\begin{enumerate}[(i)]
\item \textit{usual stochastic order} (denoted by $\mathbf{X} \leq_{st} \mathbf{Y}$) if $\mathbb{E}[\phi(\mathbf{X})] \leq \mathbb{E}[\phi(\mathbf{Y})]$, for any increasing function $\phi(\cdot)$ with finite expectations.
\item \textit{upper orthant order} (denoted by $\mathbf{X} \leq_{uo} \mathbf{Y}$) if $\bar{F}_{\mathbf{X}}(x_{1}, . . . , x_{n}) \leq \bar{F}_{\mathbf{Y}}(x_{1}, . . . , x_{n})$, for all $\mathbf{x}$, where $\bar{F}_{\mathbf{X}}$, $\bar{F}_{\mathbf{Y}}$ denote the survival function of $\mathbf{X}$ and $\mathbf{Y}$, respectively.
\item \textit{lower orthant order} (denoted by $\mathbf{X} \leq_{lo} \mathbf{Y}$) if $F_{\mathbf{X}}(x_{1}, . . . , x_{n}) \geq F_{\mathbf{Y}}(x_{1}, . . . , x_{n})$, for all $\mathbf{x}$, where $F_{\mathbf{X}}$, $F_{\mathbf{Y}}$ denote the distribution function of $\mathbf{X}$ and $\mathbf{Y}$, respectively.
\end{enumerate}
\end{Definition}

It is easy to verify that both orders,  the upper orthant and the lower orthant, are implied by the usual stochastic order.
The following stochastic order defined in \cite{laniado-lillo-romo} will be a key tool in providing some properties of the multivariate VaR that we will define in the next Section.

\begin{Definition}\label{extremOrder}
Let $\mathbf{X}$ and $\mathbf{Y}$ be two random vectors in $\mathbb{R}^{n}$, $\mathbf{X}$ is said smaller than $\mathbf{Y}$ in the extremality order in the direction $\mathbf{u}$ (denoted by $\mathbf{X}\leq_{\mathcal{E}_{\mathbf{u}}}\mathbf{Y}$) if,
\[\mathbb{P}\left[R_{\mathbf{u}}(\mathbf{X}-\mathbf{z})\geq 0\right]\leq \mathbb{P}\left[R_{\mathbf{u}}(\mathbf{Y}-\mathbf{z})\geq 0\right],\qquad \text{for all$\quad\mathbf{z}$ in $\mathbb{R}^{n}$.}\]
\end{Definition}

It is easy to show that $\mathbf{X}\leq_{\mathcal{E}_{\mathbf{u}}}\mathbf{Y}\Leftrightarrow R_{\mathbf{u}}\mathbf{X}\leq_{uo}R_{\mathbf{u}}\mathbf{Y}$. Moreover, if $\mathbf{X} \leq_{\mathcal{E}_{\mathbf{u}}} \mathbf{Y}$ then
$\mathbb{E}[\mathbf{X}]\preceq_{\mathbf{u}} \mathbb{E}[\mathbf{Y}]$, as it is proven in [\cite{laniado-lillo-romo},  \textit{Property 3.4}].
Since the multivariate VaR is based on the definition of a quantile, we also need to introduce the directional multivariate quantile given in \cite{laniado-pend}.

\begin{Definition}
Let $\mathbf{X}$ be a random vector with associated probability distribution function $\mathbb{P}$. Then the directional multivariate quantile at level $\alpha$, in direction $\mathbf{u}$ is defined as
\begin{equation}\label{qExt}
	\mathcal{Q}_{\mathbf{X}}(\alpha,\mathbf{u}):=\partial\{\mathbf{x}\in\mathbb{R}^{n}:\mathbb{P}(\mathfrak{C}_{\mathbf{x}}^{\mathbf{u}})\leq \alpha\},
\end{equation}
with $0\leq\alpha\leq 1$.
\end{Definition}

From now on, we will focus on an absolutely-continuous random vector $\mathbf{X}$ (with respect to the Lebesgue measure $\nu$ on $\mathbb{R}^{n}$) with increasing  marginal distribution functions and such that $\mathbb{E}[X_{i}] < \infty$, for $i = 1,...,n$. These conditions will be called \textit{regularity conditions}. 

\section{Directional Multivariate Value at Risk}\label{sec:MVaR}

In the univariate setting, the relationship between the quantiles related to the loss distribution  and the \textit{VaR} is obvious. In this Section, we propose a definition of multivariate \textit{VaR} for a portfolio of $n$-dependent risks, linked with the directional multivariate quantile defined in (\ref{qExt}). Besides, the output is a point in $\mathbb{R}^{n}$; that is, a vector of the same dimension as the considered portfolio of risks.
Specifically, as in the univariate case, this point defines the vertex of an oriented orthant that accumulates a probability $\alpha$, but in the direction that the investor or the risk management considers more convenient. 

\begin{Definition}\label{def:MRVaR}
Let $\mathbf{X}$ be a random vector satisfying the regularity conditions and  $0\leq \alpha \leq 1$. Then the \textit{directional multivariate Value at Risk} of $\mathbf{X}$ in direction $\mathbf{u}$ at probability level  $\alpha$ is given by
\begin{equation}\label{MRVaR}
	VaR_{\alpha}^{\mathbf{u}}(\mathbf{X})=\left(\mathcal{Q}_{\mathbf{X}}(\alpha,\mathbf{u})\bigcap\{\lambda\mathbf{u}+\mathbb{E}[\mathbf{X}]\}\right),
\end{equation}
where $\lambda\in\mathbb{R}$.
\end{Definition}

We must highlight that given a direction $\mathbf{u}$, the $VaR_{\alpha}^{\mathbf{u}}(\mathbf{X})$ is the intersection between the directional quantile at level $\alpha$, and the line defined by both the direction $\mathbf{u}$ and the mean of $\mathbf{X}$. We want to point out that the centrality tool chosen, the mean, will represent a central reference point for the random vector space, i.e., for the support of the associated probability distribution. As we will demostrate, the choice of the mean in the definition of (\ref{MRVaR}) allows us to derive desirable and interpretable analytic properties related to the risk measure.  However, other options as central reference point are possible; for example the median seen as the deepest point associated with a multivariate depth measure, which may provide a more robust risk measure (e.g. \cite{zuo-serfling,cascosDepth}).\\

\begin{figure}[htbp]
\begin{center}
\includegraphics[height=30mm,width=30mm]{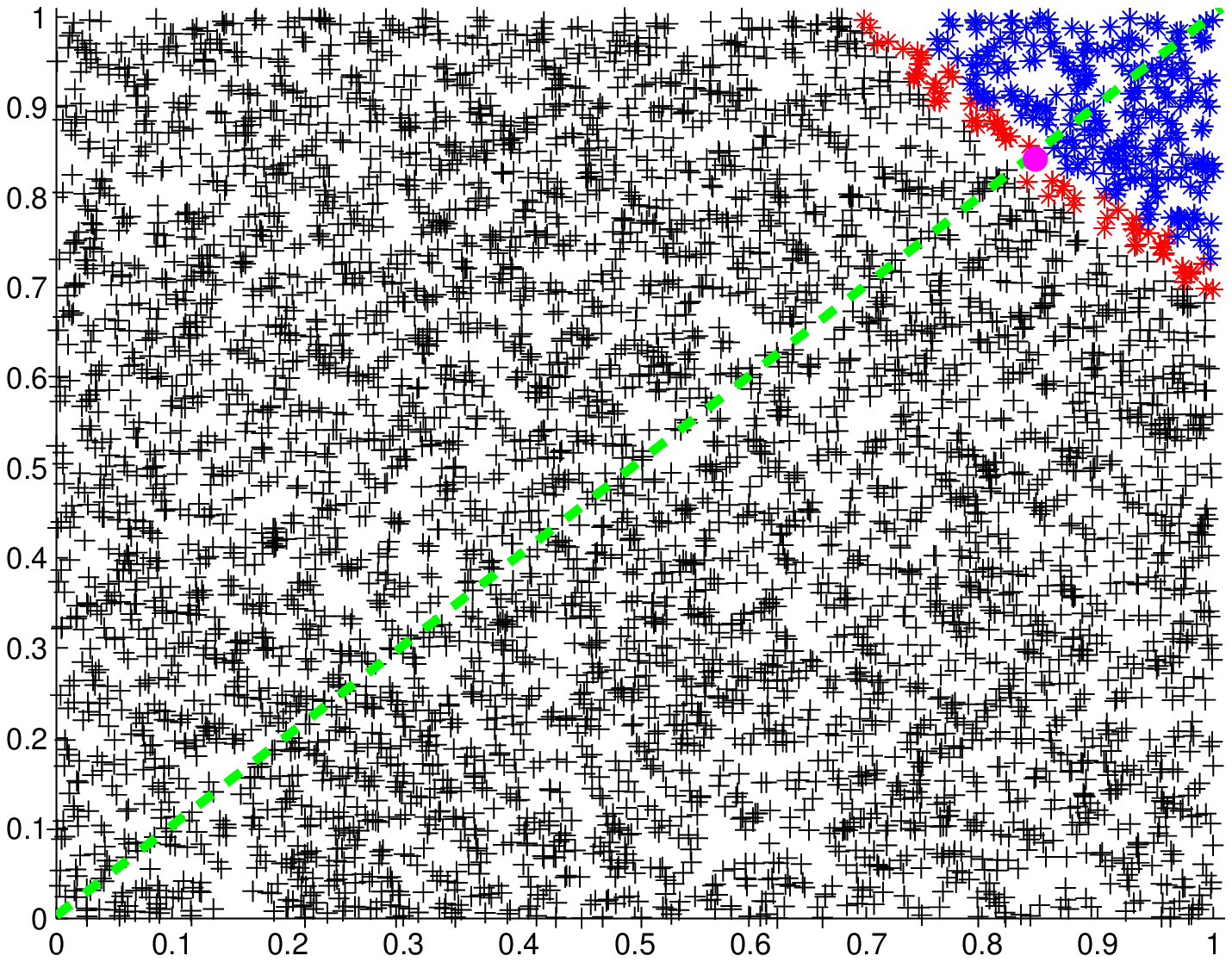}\hspace{1.4cm}\includegraphics[height=30mm,width=30mm]{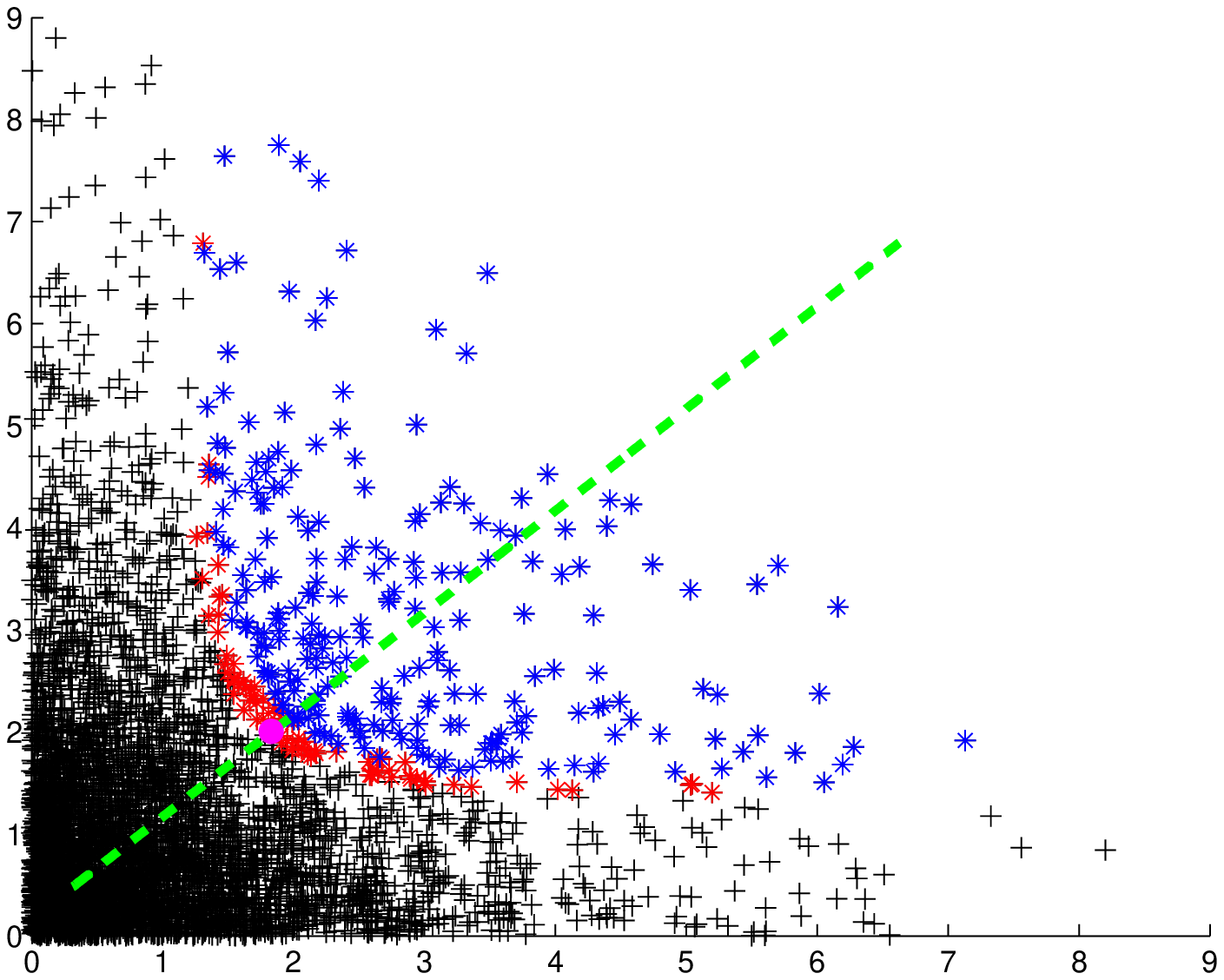}\hspace{1.3cm}\includegraphics[height=30mm,width=30mm]{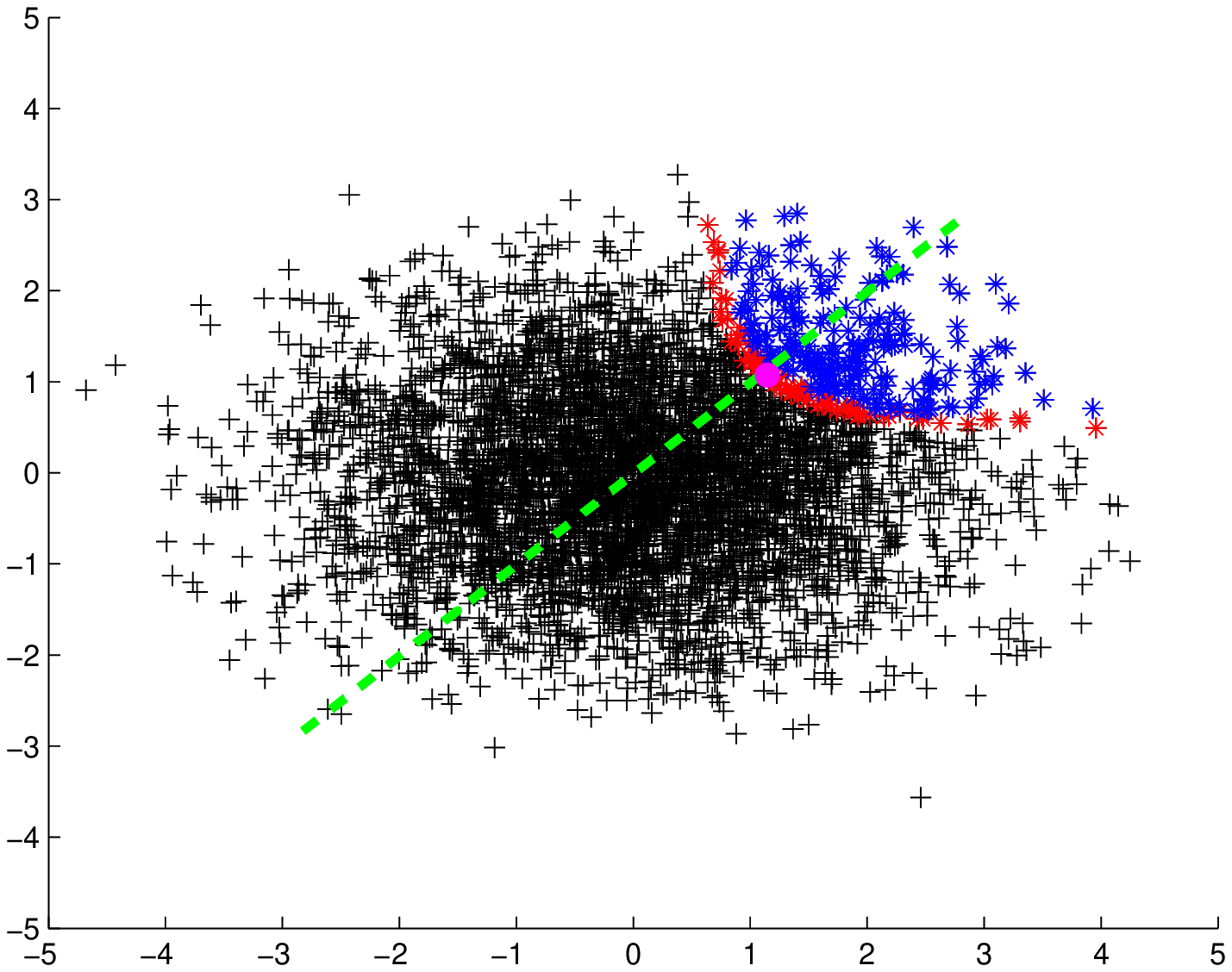}\ \ \\
\centerline{(A)\;Bivariate Uniform\hspace{1cm} (B)\;Bivariate Exponential \hspace{0.5cm} (C)\;Bivariate Normal}
    \caption{$VaR_{0.7}^{-\mathbf{e}}(\mathbf{X})$}\label{fig:MvaR1}
\end{center}
\end{figure}

\begin{figure}[htbp]
\begin{center}
\includegraphics[height=30mm,width=30mm]{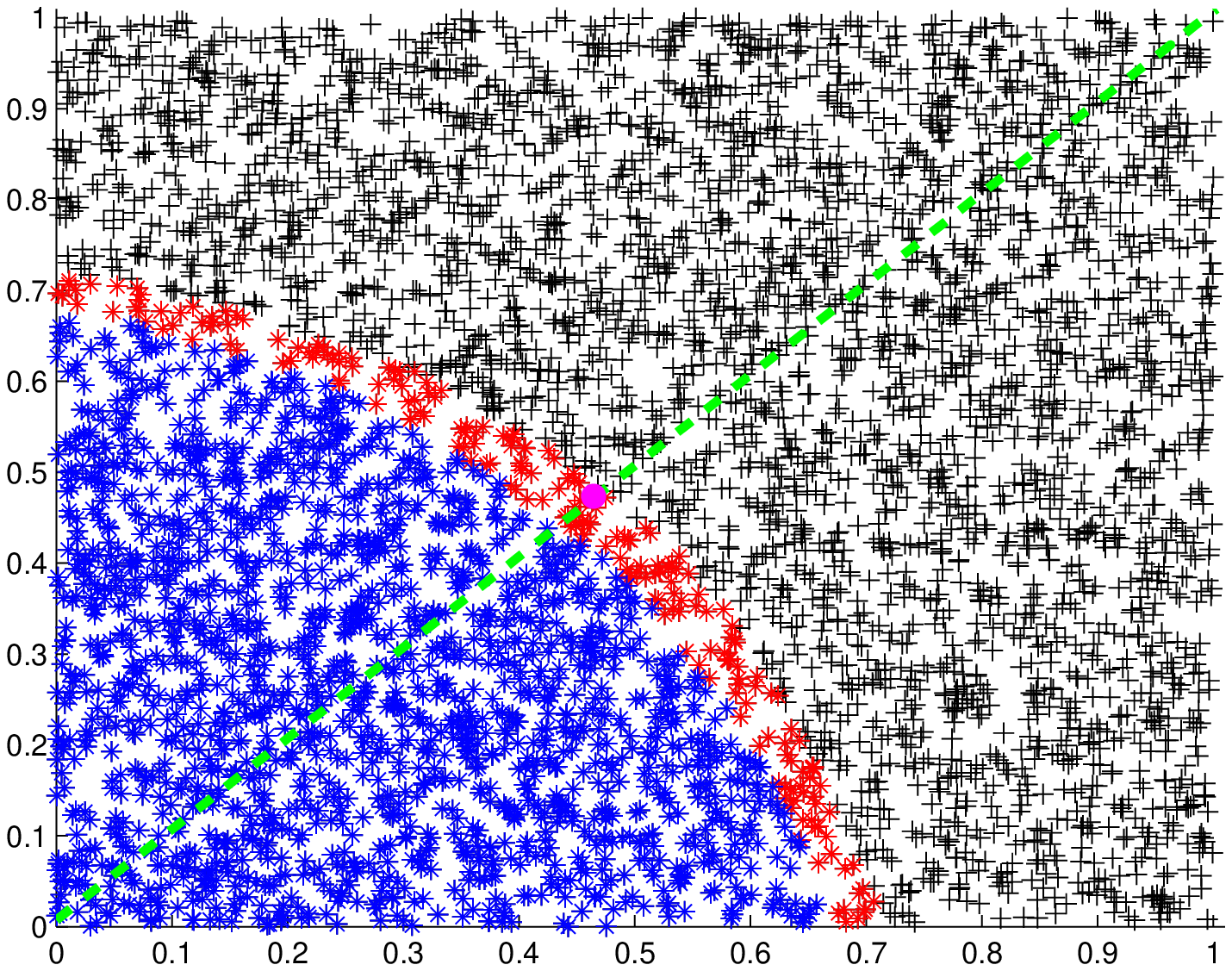}\hspace{1.4cm}\includegraphics[height=30mm,width=30mm]{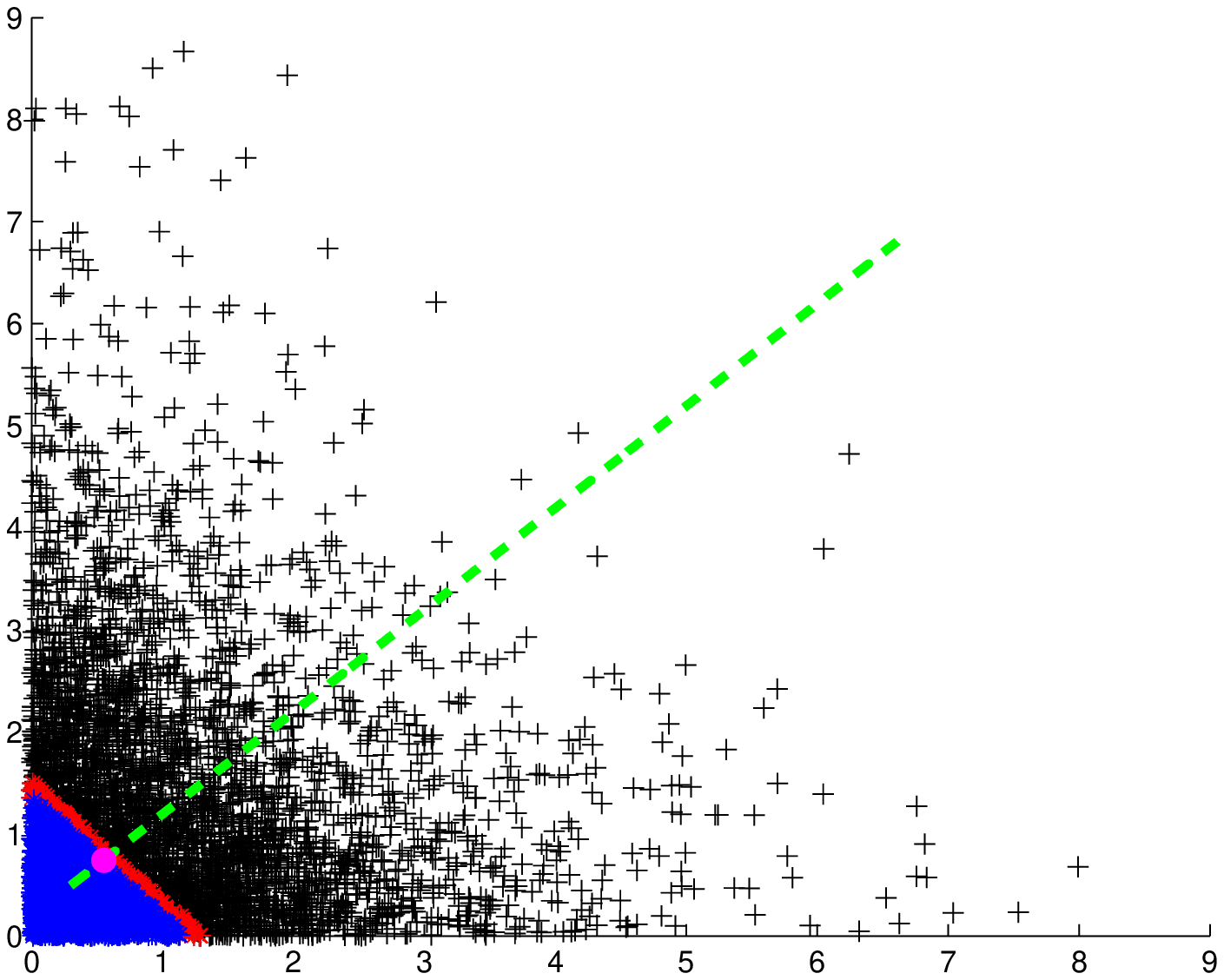}\hspace{1.3cm}\includegraphics[height=30mm,width=30mm]{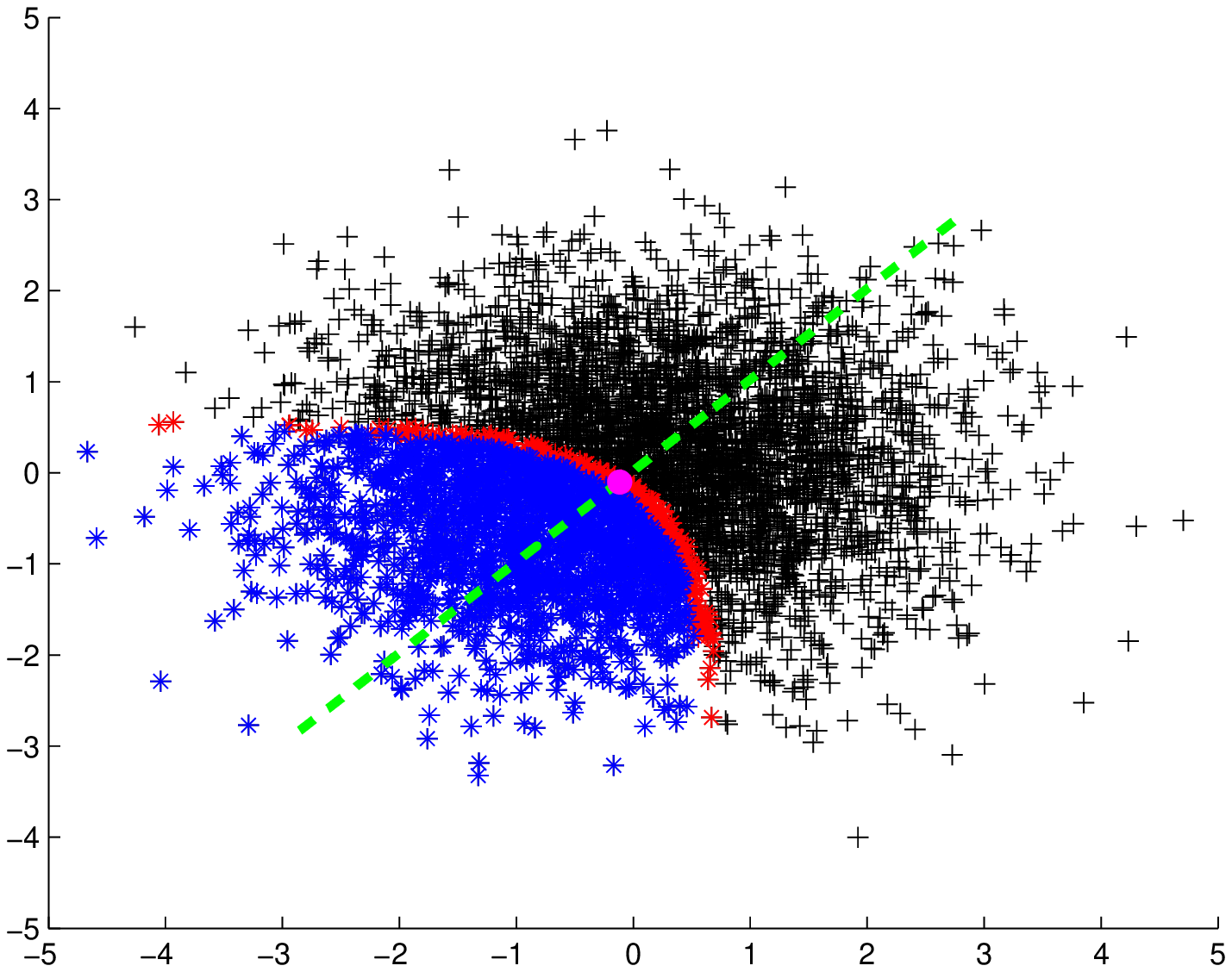}\ \ \\
\centerline{(A)\;Bivariate Uniform\hspace{1cm} (B)\;Bivariate Exponential \hspace{0.5cm} (C)\;Bivariate Normal}
    \caption{$VaR_{0.3}^{\mathbf{e}}(\mathbf{X})$}\label{fig:MvaR2}
\end{center}
\end{figure}

To illustrate this concept, you can see in Figure \ref{fig:MvaR1} some examples of the risk measure defined in (\ref{MRVaR}), for three different bivariate distributions in the direction $-\mathbf{e}$ with $\alpha=0.7$. This direction makes reference to the analysis of the distribution function of $\mathbf{X}$. Figure \ref{fig:MvaR2} presents examples with the same bivariate distributions, but in the direction $\mathbf{e}$ and for $\alpha=0.3$; that is, taking into account the information given by the survival function of $\mathbf{X}$. We call  these two directions classical directions, but the aim of this work is to show that it could be interesting to consider other directions in the analysis of risk. 

Observe that in the figures, the line in direction $\mathbf{u}$ crossing the mean in green is displayed while the quantile curve is displayed in red. The \textit{VaR} that we propose is just the intersection between the line and the quantile curve. On the other hand, the points in blue are the points "below" the level of risk $\alpha$ in the corresponding direction; meanwhile the black points are those "exceeding" the level risk. Observe Figure \ref{fig:MvaR1}, if you take any point on the blue region as a vertex of an oriented orthant in direction $-\mathbf{e}$, then the probability of that orthant will be greater than $\alpha$. It will be equal to $\alpha$ or smaller than $\alpha$ if the point is taken from the red line or black region, respectively. The same conclusion can be drawn from Figure \ref{fig:MvaR2} but in direction $\mathbf{e}$.

It is desirable that the classical  univariate \textit{VaR} agrees with our definition of \textit{VaR} in the case  $n=1$; this  fact  will be seen in the following; remember that the univariate \textit{VaR} is defined as,

\begin{equation}\label{VaR}
	VaR_{1-\alpha}(X)=\inf\{x\in\mathbb{R}: \mathbb{P}[X\geq x]\leq \alpha\},
\end{equation}

where $1-\alpha$ is usually considered closed to 1. Moreover, the \textit{VaR} may also be defined in terms of the distribution function as,

\begin{equation}\label{univariateVaR}
	VaR_{1-\alpha}(X)=\inf\{x\in\mathbb{R}: \mathbb{P}[X\leq x]\geq 1-\alpha\}.
\end{equation}

As $\mathbb{P}[X\leq x]= 1-\mathbb{P}[X\geq x]$ in the univariate setting  under regularity conditions, then (\ref{VaR}) and (\ref{univariateVaR}) are the same. To be consistent with the univariate \textit{VaR}, our definition of multivaritate \textit{VaR} agrees with the classical definition for $n=1$. That is, we have that
in terms of  $VaR_{\alpha}^{\mathbf{u}}(X)$,

\[VaR_{\alpha}^{1}(X)=VaR_{1-\alpha}(X)=VaR_{1-\alpha}^{-1}(X),\]
where $VaR_{\alpha}^{1}(X)$ is related to definition (\ref{VaR}) and $VaR_{1-\alpha}^{-1}(X)$ is related to definition (\ref{univariateVaR}).
However, this fact does not hold in the multivariate context where $F(\mathbf{x})+\bar{F}(\mathbf{x})=1$ is not true in general,  being
\begin{align}
	F(\mathbf{x})&=\mathbb{P}[\mathfrak{C}_{\mathbf{\mathbf{x}}}^{-\mathbf{e}}]=\mathbb{P}[\mathbf{X}\leq \mathbf{x}],\label{repDist}\\
	\bar{F}(\mathbf{x})&=\mathbb{P}[\mathfrak{C}_{\mathbf{x}}^{\mathbf{e}}]=\mathbb{P}[\mathbf{X}\geq \mathbf{x}].\label{repSurvival}
\end{align}

The remainder of this section is devoted to providing some properties of  $VaR_{\alpha}^{\mathbf{u}}(\mathbf{X})$ which are similar to those properties considered in the risk literature; (see \cite{artzner,burgert,cardin,rachev,cascosRM1,cascosRM2}). Specifically, we provide  properties of the multivariate $VaR_{\alpha}^{\mathbf{u}}(\mathbf{X})$ in terms of the \cite{artzner}'s properties related to  coherent risk measures in the univariate setting. Besides, we have explored other properties inherent to the multivariate response such as invariance under orthogonal transformations. All the proof for the following results is given in the Appendix.

\begin{Property}[Non-Negative Loading]\label{prop:nonNeg} If  $\lambda>0$  in (\ref{MRVaR}), then
\begin{equation}\label{eqNonNeg}
\mathbb{E}[\mathbf{X}]\preceq_{\mathbf{u}} VaR_{\alpha}^{\mathbf{u}}(\mathbf{X}).	
\end{equation}
\end{Property}

This property reflects that the risk measure  is a bound of the mean value of the losses, with respect to the partial order given in \ref{porder}.  Note that the hypothesis $\lambda>0$ is necessary, especially when $\alpha$ is chosen to be close to 0. 

\begin{Property}[Quasi-Odd Measure]\label{prop:oddMeasure} $VaR_{\alpha}^{\mathbf{u}}(\cdot)$ holds the property:
\begin{equation}
	VaR_{\alpha}^{\mathbf{u}}(-\mathbf{X})=-VaR_{\alpha}^{-\mathbf{u}}(\mathbf{X}).
\end{equation}
\end{Property}

This property shows \textit{symmetry} with respect to the random losses distribution.

\begin{Property}[Positive Homogeneity and Translation Invariance]\label{prop:homTrans} Let $c\in \mathbb{R}_{+}$, $\mathbf{b}\in\mathbb{R}^{n}$ and $\mathbf{Y}=c\mathbf{X} + \mathbf{b}$, then,
\begin{equation}
	VaR_{\alpha}^{\mathbf{u}}(\mathbf{Y})=c VaR_{\alpha}^{\mathbf{u}}(\mathbf{X})+\mathbf{b}.
\end{equation}
\end{Property}

\begin{Property}[Consistency w.r.t. extremality stochastic order]\label{prop:consistencia} Let $\mathbf{X}$ and $\mathbf{Y}$ be random vectors satisfying the \textit{regularity conditions}. If $\mathbb{E}[\mathbf{Y}]=c\mathbf{u}+\mathbb{E}[\mathbf{X}]$ with $c>0$, and $\mathbf{X} \leq_{\mathcal{E}_{u}} \mathbf{Y}$, then:
\begin{equation}\label{eqMonot}
	VaR_{\alpha}^{\mathbf{u}}(\mathbf{X})\preceq_{u} VaR_{\alpha}^{\mathbf{u}}(\mathbf{Y}).
\end{equation}
\end{Property}

\begin{Property}[Orthogonal Quasi-Invariance]\label{prop:rot}
Let $Q$ be an orthogonal transformation. Then,
\begin{equation}
	VaR_{\alpha}^{Q\mathbf{u}}(Q\mathbf{X})=QVaR_{\alpha}^{\mathbf{u}}(\mathbf{X}).
\end{equation}
\end{Property}

\begin{Property}[Non-Excessive Loading]\label{prop:nonExc} Let $R_{\mathbf{u}}$ be the orthogonal matrix described in (\ref{convexcone}). Then,
\begin{equation}\label{eqNonExc}
VaR_{\alpha}^{\mathbf{u}}(\mathbf{X})\preceq_{\mathbf{u}} R_{\mathbf{u}}'\sup_{\omega\in\Omega}\{R_{\mathbf{u}} \mathbf{X}(\omega)\}.
\end{equation}
\end{Property}

This property shows that $VaR_{\alpha}^{\mathbf{u}}(\mathbf{X})$ is upper bounded by the supreme of the losses in the direction considered.
Another good property which is desirable in the literature for risk measures is the subadditivity. As is well-known, the classical univariate \textit{VaR}  is not  a subadditivity measure. However,  there are conditions that ensure the tail region subadditivity property (see \cite{artzner,heyde,danielson}). In the same way, we highlighted that the $VaR_{\alpha}^{\mathbf{u}}(\mathbf{X})$ is not subadditive in general, but we will prove that this property holds under some conditions. A previous definition is necessary.

\begin{Definition}
A random vector $\mathbf{X}$ has regularity varying, with tail index $\beta$ if there is a function $\phi(t)>0$ that is regularly varying at infinity with exponent $\frac{1}{\beta}$ and a non-zero measure $\mu(\cdot)$ on the Borel $\sigma-$field $\mathcal{B}([0,\infty]^{n}\backslash\{\mathbf{0}\})$ such that,
\begin{equation}
t\mathbb{P}[(\phi(t))^{-1}\mathbf{X}\in \cdot]\stackrel{v}{\rightarrow} \mu(\cdot),
\end{equation}
when $t\rightarrow\infty$ (see \cite{jessen,resnick1}).
\end{Definition}

In this case, the measure has the property
\begin{equation}\label{regVarying}
\mu(cB)=c^{-\beta}\mu(B),
\end{equation}
for any $c>0$ and $B$ a Borel set.

With this definition, we can state the tail region subadditivity property of the  $VaR_{\alpha}^{\mathbf{u}}(\cdot)$.

\begin{Property}[Tail Region Subadditivity]\label{prop:sub} Let $\mathbf{X}$ and $\mathbf{Y}$ be random vectors, with the same mean $\mathbf{m}$. If $(\mathbf{X},\mathbf{Y})$ is a regularly varying random vector with index $\beta>1$ and non-degenerate tails then, the $VaR_{\alpha}^{\mathbf{u}}(\cdot)$ is subadditive in the tail region in direction $\mathbf{u}=\frac{\mathbf{m}}{||\mathbf{m}||}$, i.e.,
\begin{equation}\label{tailSubadd}
	VaR_{\alpha}^{\mathbf{u}}(\mathbf{X}+\mathbf{Y})\preceq_{\mathbf{u}} VaR_{\alpha}^{\mathbf{u}}(\mathbf{X})+VaR_{\alpha}^{\mathbf{u}}(\mathbf{Y}).
\end{equation}
\end{Property}

Note that the Property \ref{prop:sub} could be extended to random vectors with means satisfying $\mathbb{E}[\mathbf{X}]=c\mathbb{E}[\mathbf{Y}]$ for $c>0$. As you can see, the property ensures  that at least in the direction of the mean loss, it is useful to merge  two risky activities in order to diversify the risk.

\section{Comparison of the univariate \textit{VaR} componentwise   and the Directional Multivariate \textit{VaR}}\label{sec:comp}

The aim of this section is to compare the components of $VaR_{\alpha}^{\mathbf{u}}(\mathbf{X})$ with the univariate \textit{VaR} related to  each  marginal distribution of $\mathbf{X}$.
But prior to this we need to remember the definition of a multivariate quasi-concave function.

\begin{Definition}\label{def:quasiConcave}
A multivariate function $g:\mathbb{R}^{n}\rightarrow\mathbb{R}$ is a quasi-concave function if the upper-level set $U_{q}:=\{\mathbf{x}\in\mathbb{R}^{n} : g(x)\geq q\}$ is a convex set for all $q\in\mathbb{R}$. Or equivalently, the complementary of the lower set $L_{q}:=\{\mathbf{x}\in\mathbb{R}^{n} : g(x)\leq q\}$ is a convex set for all $q\in\mathbb{R}$.
\end{Definition}

We want to point out that both the  distribution and  survival functions, in general, satisfy Definition \ref{def:quasiConcave}. This fact was proved in \cite{tibiletti-quasi}, and therefore it is not a restrictive condition for the functions considered in this paper. Let us denote by $X_{i}$ the $i$-th marginal of the random vector $\mathbf{X}$ and by $[\cdot]_{i}$ the $i$-th component related to a point in $\mathbb{R}^{n}$.
The following result provides comparisons between the components of the  multivariate \textit{VaR} introduced in this work and the classical univariate \textit{VaR}.

\begin{Proposition}\label{prop:relMarginal} Consider a random vector $\mathbf{X}$ satisfying the regularity conditions. Assume that its survival function $\bar{F}$ is quasi-concave. Then, for all $\alpha\in (0, 1)$:
\[VaR_{1-\alpha}(X_{i})\geq \left[VaR_{\alpha}^{\mathbf{e}}(\mathbf{X})\right]_{i},\qquad \text{ for all}\quad i=1,...,n.\]
Moreover, if its multivariate distribution function $F$ is quasi-concave, then, for all $\alpha\in (0, 1)$, we have that
\[\left[VaR_{1-\alpha}^{-\mathbf{e}}(\mathbf{X})\right]_{i}\geq VaR_{1-\alpha}(X_{i}),\qquad\text{ for all}\quad i=1,...,n.\]
\end{Proposition}

The proof is given in the Appendix. As you can see, the preceding result can be extended considering other directions as follows.
\begin{Corollary}
Let  $\mathbf{X}$ be a random variable satisfying the regularity conditions and fix a direction $\mathbf{u}$. If the survival function of $R_{\mathbf{u}}\mathbf{X}$ is a quasi-concave function,
then,  for all $0 \leq \alpha \leq 1$,
\[VaR_{1-\alpha}([R_{\mathbf{u}}X]_{i})\geq \left[R_{\mathbf{u}}VaR_{\alpha}^{\mathbf{u}}(\mathbf{X})\right]_{i},\qquad \text{ for all}\quad i=1,...,n.\]
Besides, if $R_{\mathbf{u}}X$ has a quasi-concavity cumulative distribution, then
\[\left[R_{\mathbf{u}}VaR_{1-\alpha}^{-\mathbf{u}}(\mathbf{X})\right]_{i}\geq VaR_{1-\alpha}([R_{\mathbf{u}}X]_{i}),\qquad\text{ for all}\quad i=1,...,n,\]
where $R_{\mathbf{u}}$ is the orthogonal transformation defined in (\ref{convexcone}).
\end{Corollary}

The proof is straightforward from Proposition \ref{prop:rot} and Proposition \ref{prop:relMarginal}.
Therefore, by linking the previous results we have the following inequality for all pairs $(\mathbf{u} ,\alpha)$, $(-\mathbf{u} ,1-\alpha)$.

\begin{equation}\label{uplowVaR}
	VaR_{\alpha}^{\mathbf{u}}(\mathbf{X})\preceq_{\mathbf{u}} VaR_{1-\alpha}^{-\mathbf{u}}(\mathbf{X}).
\end{equation}


This relationship allows us to define a   \textit{directional upper VaR} and a \textit{directional lower VaR} in a similar way to \cite{ep} and \cite{bernardino}, but with a unified notation. Specifically, we have introduced the following definitions:

The \textit{upper VaR in direction} $\mathbf{u}$ is,

\begin{equation}\label{eq:upVaR}
\overline{VaR}_{\alpha}^{\mathbf{u}}(\mathbf{X})=VaR_{\alpha}^{\mathbf{u}}(\mathbf{X}),
\end{equation}

The \textit{lower VaR in a direction} $\mathbf{u}$ is,

\begin{equation}\label{eq:lowVaR}
\underline{VaR}_{\alpha}^{\mathbf{u}}(\mathbf{X})=VaR_{1-\alpha}^{-\mathbf{u}}(\mathbf{X}).
\end{equation}

\begin{figure}[htbp]
\begin{center}
	\includegraphics[height=60mm,width=60mm]{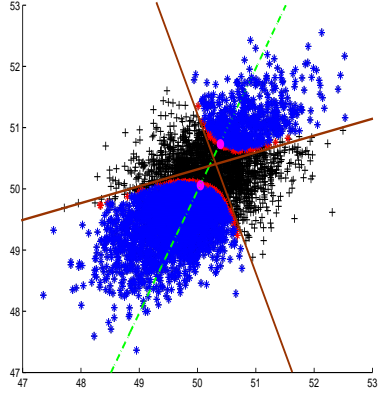}
	\caption{Lower and upper $VaR_{\alpha}^{\mathbf{u}}(\mathbf{X})$ with $\mathbf{u}=(\frac{1}{\sqrt{5}},\frac{2}{\sqrt{5}})$ and $\alpha=0.3$ for a bivariate Normal.}
\label{fig:lowup}
\end{center}
\end{figure}

An example of these concepts is displayed in Figure \ref{fig:lowup}, where we can see in a bivariate normal distribution, the \textit{upper VaR in direction } $\mathbf{u}=(\frac{1}{\sqrt{5}},\frac{2}{\sqrt{5}})$ for a level of risk $\alpha=0.3$, and the corresponding \textit{lower VaR in direction } $-\mathbf{u}$ and level risk $1-\alpha$. 
Note that we can describe in the plot types of asymptotes for the quantile curves, furthermore these asymptotes will be the univariate quantiles for each marginal of the rotated random vector $R_{\mathbf{u}}\mathbf{X}$ at the same $\alpha$, where the rotation matrix $R_{\mathbf{u}}$ is the same as in (\ref{convexcone}). These asymptotes can be seen as a generalization of those defined in \cite{bcsll} for the quantile curves in the classical directions.

Another practical situation where the link between the multivariate \textit{VaR} and the univariate \textit{VaR} is interesting (see e.g. \cite{ep,wang1,bernard}), is when it is necessary to give bounds of the univariate \textit{VaR} over a linear transformation of the marginal losses; for instance, when the transformation by the portfolio weights vector is considered, i.e., when the objective random variable is

\[Z = \mathbf{w}'\mathbf{X},\]

where $\mathbf{w}$ is the  vector of the portfolio weights. Since it is difficult to obtain the  \textit{VaR} of $Z$ mainly when the components of the portfolio are not independent, there is special interest in obtaining at least a  bound for $VaR_{\alpha}(Z)$. Fortunately, we can give an upper-bound using our directional approach.

\begin{Proposition}\label{prop:cotaHenry}
Let $\mathbf{u} = -\frac{\mathbf{w}}{||\mathbf{w}||}$ be the unitary vector in direction of the portfolio weights. If $\mathbf{x}\in \mathcal{Q}_{\mathbf{X}}(\alpha,\mathbf{u})$, then $\mathbf{w}'\mathbf{x} \geq VaR_{\alpha}(Z)$.
\end{Proposition}

The proof is given in the Appendix.

Specifically as a consequence of Proposition \ref{prop:cotaHenry}, we have that
\begin{equation}\label{eq:cotaHenry}
\mathbf{w}'VaR_{\alpha}^{-\frac{\mathbf{w}}{||\mathbf{w}||}}(\mathbf{X}) \geq VaR_{\alpha}(Z).
\end{equation}

This result is another justification to consider a directional approach of the multivariate \textit{VaR}, as well as its utility in financial applications.

\section{Directional multivariate \textit{VaR} and copulas}\label{sec:cop}

Researchers refer to copulas as "the multivariate distribution functions whose one-dimensional marginal distributions are uniform in $[0,1]$". For an extensive discussion of copulas, we refer the reader to  \cite{nelsen}.
This powerful tool allows the definition of scale-free measures of dependence and families of multivariate distributions. Two aspects are important in multivariate distributions, the distribution of the marginals and the dependence structure among them. The concept of copula fully describes  the overall  structure of dependence between the marginal variables and provides a global model for their stochastic behavior. The important result that links these two aspects is Sklar's theorem that allows, in terms of a copula, to write the multivariate distribution function as,
\begin{equation}\label{relcopula}
	F(x_{1},\cdots,x_{n})=C(F_{1}(x_{1}),\cdots,F_{n}(x_{n})),
\end{equation}

where $F$ is the join distribution function, $F_{1},..., F_{n}$ its marginals distribution and $C$ the copula, which according to  Sklar's theorem always exists.
 The copulas become a powerful tool to  find   closed expression of multivariate quantiles for  special families of copulas.  For example, in   finance when the losses are modeled in percentage  terms, it is of practical importance to find closed expressions for the risk measures expressed in terms of the copula since the support of the losses will be the unitary hyper cube of dimension $n$.

Hence, the objective of this section is to analyze how the  $VaR_{\alpha}^{\mathbf{u}}(\mathbf{X})$ can be obtained in terms of some
 families of copulas. The first result shows the representation of the  $VaR_{\alpha}^{\mathbf{u}}(\mathbf{X})$ restricted to bivariate  copulas.
Let $\mathbf{X}$ be a bivariate random vector with marginals uniformly distributed in the interval $[0,1]$. In this case, the distribution function of  $\mathbf{X}$ is a copula with density $c(\cdot,\cdot)$.
 It is well known that  $E\left[\mathbf{X}\right]= (\frac{1}{2},\frac{1}{2})$. Note that assuming $n=2$, a direction $\mathbf{u}=(u_{1},u_{2})$ can be characterized by a angle $\theta$
 such that $\tan \theta=u_{2}/u_{1}$, and then,  $\mathbf{u}=(\cos \theta, \sin \theta)$. Following with the notation given by the angles,  the $VaR_{\alpha}^{\mathbf{u}}(\mathbf{X})$ must be a point on the line $l_{\theta}$ defined by,

\begin{equation}\label{lineCop}
	l_{\theta}:=\begin{cases}
	\left\{(w_{1},w_{2}) : w_{2}=\frac{w_{1} \sin(\theta)-\frac{1}{2}(\sin(\theta)-\cos(\theta))}{\cos(\theta)}\right\},\qquad &\text{ if }\quad \cos(\theta)\neq 0,\\
	\left\{(w_{1},w_{2}) : w_{1}\in[0,1], w_{2}=\frac{1}{2}\right\},\qquad &\text{ if }\quad \cos(\theta)=0.
	\end{cases}
\end{equation}

Therefore, given a   direction $\theta$,  $VaR_{\alpha}^{\mathbf{u}}(\mathbf{X})$ is characterized by its first component and the second one is obtained using (\ref{lineCop}). Now,  the first component can be obtained by solving the following integral equation,

\begin{equation}\label{copVaR}
	\int\int_{D_{\theta}(w_{1})}{c(s,t)dtds}=\alpha,
\end{equation}

where  $D_{\theta}(w_{1})$ is given by the intersection of the unitary square $[0,1]\times[0,1]$ and the oriented quadrant with direction determined by $\theta$ and vertex  $(w_{1},l_{\theta}(w_{1}))$. Specifically,
$D_{\theta}(w_{1})$  can be expressed  in terms of the unknown $w_{1}$ by using the semi-lines $l^{1}_{\theta}(w_{1})$, $l^{2}_{\theta}(w_{1})$ that bound the corresponding quadrant which are defined as,

\scriptsize
\[\begin{aligned}
&l^{1}_{\theta}(w_{1}):=\\
&\left\{(z_{1},z_{2}) : z_{2}\cos\left(\theta-\frac{\pi}{4}\right)-z_{1}\sin\left(\theta-\frac{\pi}{4}\right)=w_{1}\left(\tan(\theta)\cos\left(\theta-\frac{\pi}{4}\right)-\sin\left(\theta-\frac{\pi}{4}\right)\right)-\frac{1}{2}\left(\tan(\theta)-1\right)\cos\left(\theta-\frac{\pi}{4}\right)\right\}\\
&l^{2}_{\theta}(w_{1}):=\\
&\left\{(z_{1},z_{2}) : z_{2}\sin\left(\theta-\frac{\pi}{4}\right)+z_{1}\cos\left(\theta-\frac{\pi}{4}\right)=w_{1}\left(\tan(\theta)\sin\left(\theta-\frac{\pi}{4}\right)+\cos\left(\theta-\frac{\pi}{4}\right)\right)-\frac{1}{2}(\tan(\theta)-1)\sin\left(\theta-\frac{\pi}{4}\right)\right\}\\
\end{aligned}\]

\normalsize
For instance, if $\theta\in (\frac{\pi}{4},\frac{\pi}{2})$,  we can write the integral equation as follows:

\begin{equation}\label{copBivariada}
\int_{\min\{l^{2}_{\theta}(w_{1})\bigcap \{z_{1}=0\},0\}}^{w_{1}}\int_{l^{2}_{\theta}(w_{1})}^{1}{c(s,t)dtds}+\int_{w_{1}}^{\min\{l^{1}_{\theta}(w_{1})\bigcap \{z_{1}=1\},1\}}\int_{l^{1}_{\theta}(w_{1})}^{1}{c(s,t)dtds}=\alpha.
\end{equation}

\begin{figure}[htbp]
\begin{center}
	\includegraphics[height=60mm,width=60mm]{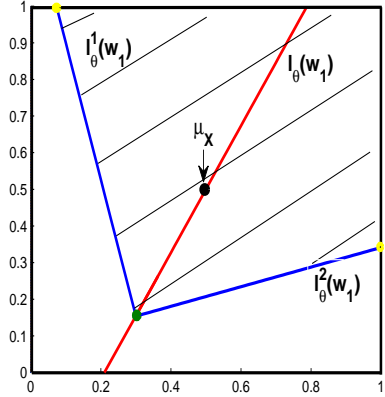}
	\caption{Quadrant given by $\theta\in (\frac{\pi}{4},\frac{\pi}{2})$ and vertex over the line $l_{\theta}$.}
	\label{fig:orthantEj}
\end{center}
\end{figure}

Figure \ref{fig:orthantEj} shows a case of the  region $D_{\theta}(w_{1})$ with  $\theta\in (\frac{\pi}{4},\frac{\pi}{2})$ being the solution to  (\ref{copBivariada}), a point over the line $l_{\theta}$.
In summary, we can obtain  $VaR_{\alpha}^{\mathbf{u}}(\mathbf{X})$ for a given bivariate vector with copula density $c(\cdot,\cdot)$.

Now, we will focus on the Archimedean family of copulas broadly used in the literature whose definition is the following:

\normalsize

\begin{Definition}[Archimedean Copulas]
Let $\phi:[0,1]\rightarrow [0,\infty)$ be a continuous, convex and strictly decreasing function with $\phi(\mathbf{1})=0$. Let $\phi^{-1}(\cdot)$ be a pseudo-inverse function of $\phi(\cdot)$. Then an Archimedean copula $C(v_{1},\cdots,v_{n})$ is defined by

\begin{equation}\label{arqCopula}
	C(v_{1},\cdots,v_{n})=\phi^{-1}(\phi(v_{1})+\cdots+\phi(v_{n})).
\end{equation}
\end{Definition}

In this case, for  an $n$-dimensional random variable with distribution function as belonging to  the Archimedean family of copulas with generator $\phi(\cdot)$,  $VaR_{\alpha}^{-\mathbf{e}}(\mathbf{X})$ is given by the vector with all components equal to

\begin{equation}\label{copArqMRVaR}
	[VaR_{1-\alpha}^{-\mathbf{e}}(\mathbf{X})]_{i} = \phi^{-1}\left(\frac{\phi(1-\alpha)}{n}\right).
\end{equation}

Moreover, if $\mathbf{X}$ has a survival copula $\breve{C}$ belonging to the Archimedean family with generator $\breve{\phi}(\cdot)$,
the equivalent Sklar's representation gives the relation $\bar{F}_{\mathbf{X}}(x_{1},\cdots,x_{n})=\breve{C}(\bar{F}_{1}(x_{1}),\cdots,\bar{F}_{n}(x_{n}))$, where $\bar{F}$ is the join survival function and $\bar{F}_{1},..., \bar{F}_{n}$ its marginal survival functions. Hence, we obtain that:

\begin{equation}\label{surArqMRVaR}
	[VaR_{\alpha}^{\mathbf{e}}(\mathbf{X})]_{i} = 1-\breve{\phi}^{-1}\left(\frac{\breve{\phi}(\alpha)}{n}\right).
\end{equation}

Remember that if a vector $\mathbf{X}$ has a copula $C$, then the survival copula of $\mathbf{1-X}$ will also be  $C$. Therefore, if $\mathbf{X}\stackrel{\text{d}}{=}\mathbf{1-X}$,
then the copula of  $\mathbf{X}$ and its survival copula are the same; for example, Frank's copula in the Archimedean family holds this property as well as the elliptical family of copulas. Then, in this case the closed expression for $VaR_{\alpha}^{\mathbf{e}}(\mathbf{X})$ is the reflection point of $VaR_{1-\alpha}^{-\mathbf{e}}(\mathbf{X})$ with respect to the point $(\frac{1}{2},\cdots,\frac{1}{2})$.

Now we will present some examples using some Archimedean copulas. Firstly, we are going to use Frank's subclass to present an example of $VaR_{\alpha}^{\mathbf{u}}(\mathbf{X})$ for any direction $\mathbf{u}$ in the bivariate case.
Later we will present some comparisons between the \textit{lower orthant VaR} $\equiv \underline{VaR}_{\alpha}(\mathbf{X})$ and the \textit{upper orthant VaR} $\equiv \overline{VaR}_{\alpha}(\mathbf{X})$ developed by \cite{bernardino} with the $VaR_{\alpha}^{\mathbf{u}}(\mathbf{X})$ but considering a  $n$-dimensional  copula belonging to Clayton's subclass. Let's define these two subclasses.

\begin{enumerate}[(i)]

\item  \textbf{Frank Copula}: The generated function of this copula is
\begin{equation}\label{genFrank}
	\phi_{\beta}(r)=-ln\left(\frac{e^{-\beta r}-1}{e^{-\beta}-1}\right)\quad \text{ and }\quad \phi^{-1}_{\beta}(s)=-\frac{1}{\beta}ln(1-(1-e^{-\beta})e^{-s}),
\end{equation}

\begin{align}\label{frank}
	C_{\beta}(v_{1},v_{2})&=-\frac{1}{\beta}ln\left(1+\frac{(e^{-\beta v_{1}}-1)(e^{-\beta v_{2}}-1)}{e^{-\beta}-1}\right),\\
	c_{\beta}(v_{1},v_{2})&=-\frac{\beta(1-e^{-\beta})e^{-\beta(v_{1}+v_{2})}}{\left((e^{-\beta v_{1}}-1)(e^{-\beta v_{2}}-1)-(e^{-\beta}-1)\right)^{2}},
\end{align}
where $\beta\in\mathbb{R}\backslash \{0\}$.

\item \textbf{Clayton Copula}: This family is generated by

\begin{equation}\label{genClayton}
	\phi_{\beta}(r)=\frac{1}{\beta}(r^{-\beta}-1)\quad \text{ and }\quad \phi^{-1}_{\beta}(s)=(1+\beta s)^{-1/\beta},
\end{equation}

\begin{equation}\label{clayton}
	C_{\beta}(v_{1},v_{2})=\max\left\{(v_{1}^{-\beta}+v_{2}^{-\beta}-1)^{1/\beta},0\right\},
\end{equation}
where $\beta\in [-1,0)\cup (0.+\infty]$.
\end{enumerate}

\begin{figure}[htbp]
\includegraphics[height=45mm,width=45mm]{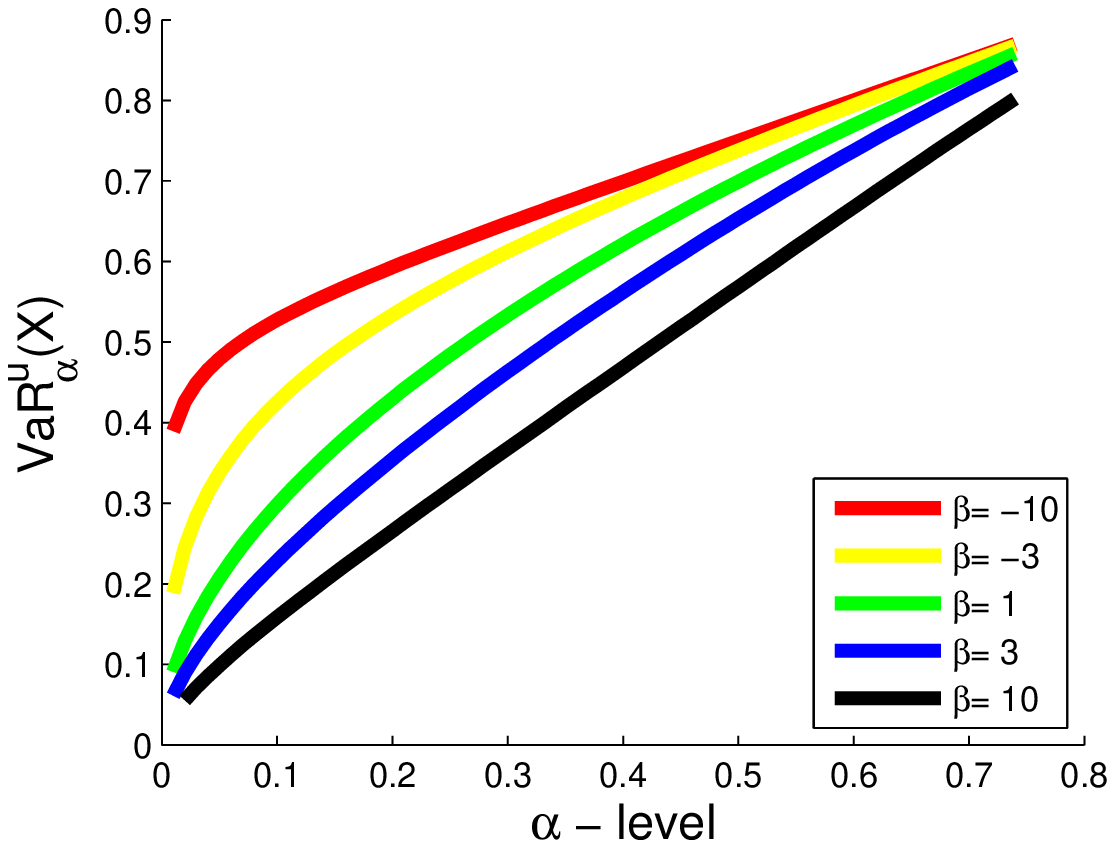}\hspace{3cm}\includegraphics[height=45mm,width=45mm]{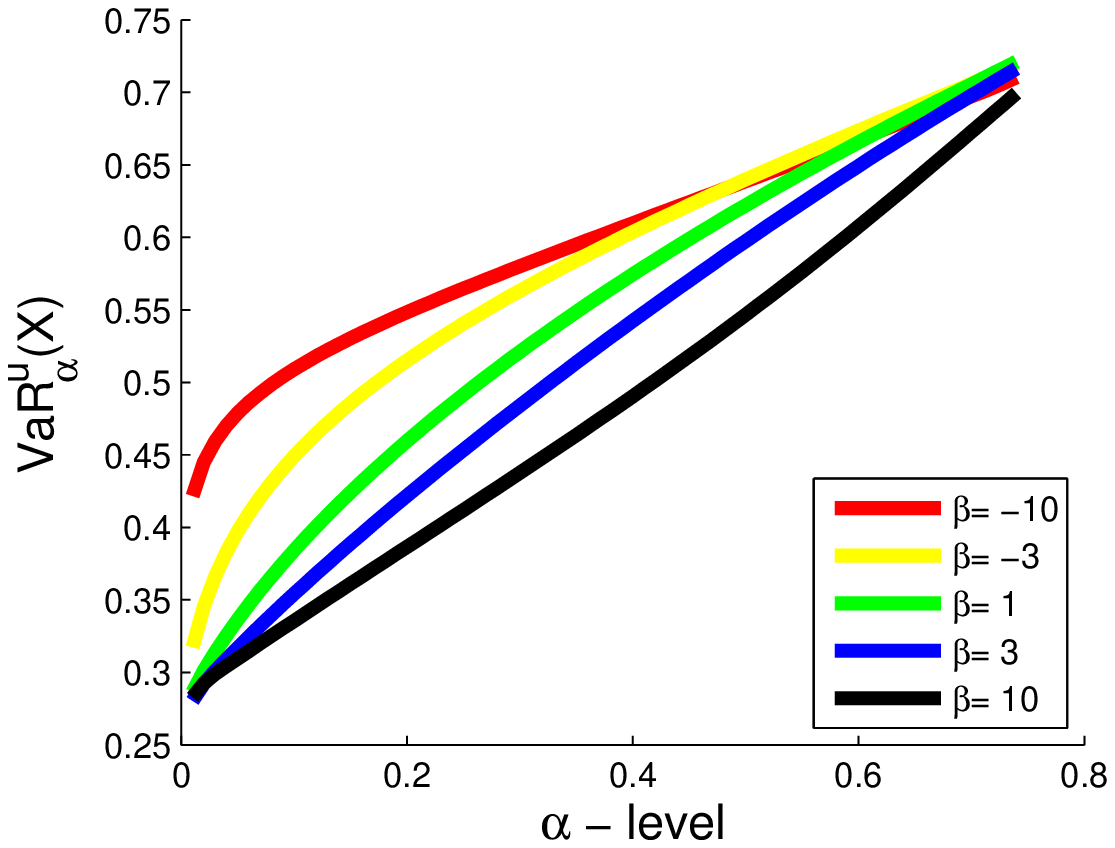}\ \ \\
\centerline{a)\;\;Direction $-\mathbf{e}=-\frac{\sqrt{2}}{2}[1,1]'$\hspace{3cm}b)\;\;Direction $\mathbf{u}=-\frac{3\sqrt{5}}{5}[\frac{1}{3},\frac{2}{3}]'$}
    \caption{Behavior for the first component in   $VaR_{\alpha}^{\mathbf{u}}(\mathbf{X})$ varying $\alpha$.}\label{fig:MRVaRcop}
\end{figure}

In Figure \ref{fig:MRVaRcop} we have drawn the first component of the directional $VaR_{\alpha}^{\mathbf{u}}(\mathbf{X})$ for a bivariate random vector, with density given by the Frank copula density. The left plot is related to  $\mathbf{u}=-\mathbf{e}$ and the right plot is related to  $\mathbf{u}=-\frac{1}{\sqrt{5}}(1,2)$. Both plots present the changes as $0 \leq \alpha \leq 1$ for different values of the dependence parameter $\beta$.

We can see in Figure \ref{fig:MRVaRcop}  the  dependence of $VaR_{\alpha}^{\mathbf{u}}(\mathbf{X})$ with respect to $\beta$. Note that  as $\beta\rightarrow =\pm \infty$ and the direction  is $\pm\mathbf{e}$, we will get  the extreme cases known as comonotonic and counter-monotonic, respectively. In the left plot, it  can be seen that the comonotonic case matches  with the vector composed of the univariate \textit{VaR} on the marginals, which in this case is given by the vector $[VaR_{\alpha}^{-\mathbf{e}}(\mathbf{X})]_{i}$.  In addition, it is well known that rotations over random vectors do not preserve the dependence structure in the rotated distribution; furthermore, this fact is captured in the right plot where the change of direction shows the rotations of the measure in each dependence parameter considered.

Let $\mathbf{X}$ be a random vector with distribution function belonging to the Clayton copula subclass. Hence $\mathbf{1-X}$ is a random vector with Clayton survival copula. We have presented the comparison of the first component of $VaR_{\alpha}^{-\mathbf{e}}(\mathbf{X})$ with $\underline{VaR}_{\alpha}(\mathbf{X})=\mathbb{E}\left[\mathbf{X} | F(\mathbf{x})=\alpha\right]$ and  $VaR_{\alpha}^{\mathbf{e}}(\mathbf{1-X})$ with $\overline{VaR}_{\alpha}(\mathbf{1-X})=\mathbb{E}\left[\mathbf{X} | \bar{F}(\mathbf{x})=1-\alpha\right]$, the correspondent \textit{lower orthant VaR} and \textit{upper orthant VaR} developed by \cite{bernardino}.

\begin{figure}[htbp]
\begin{center}
\includegraphics[height=45mm,width=60mm]{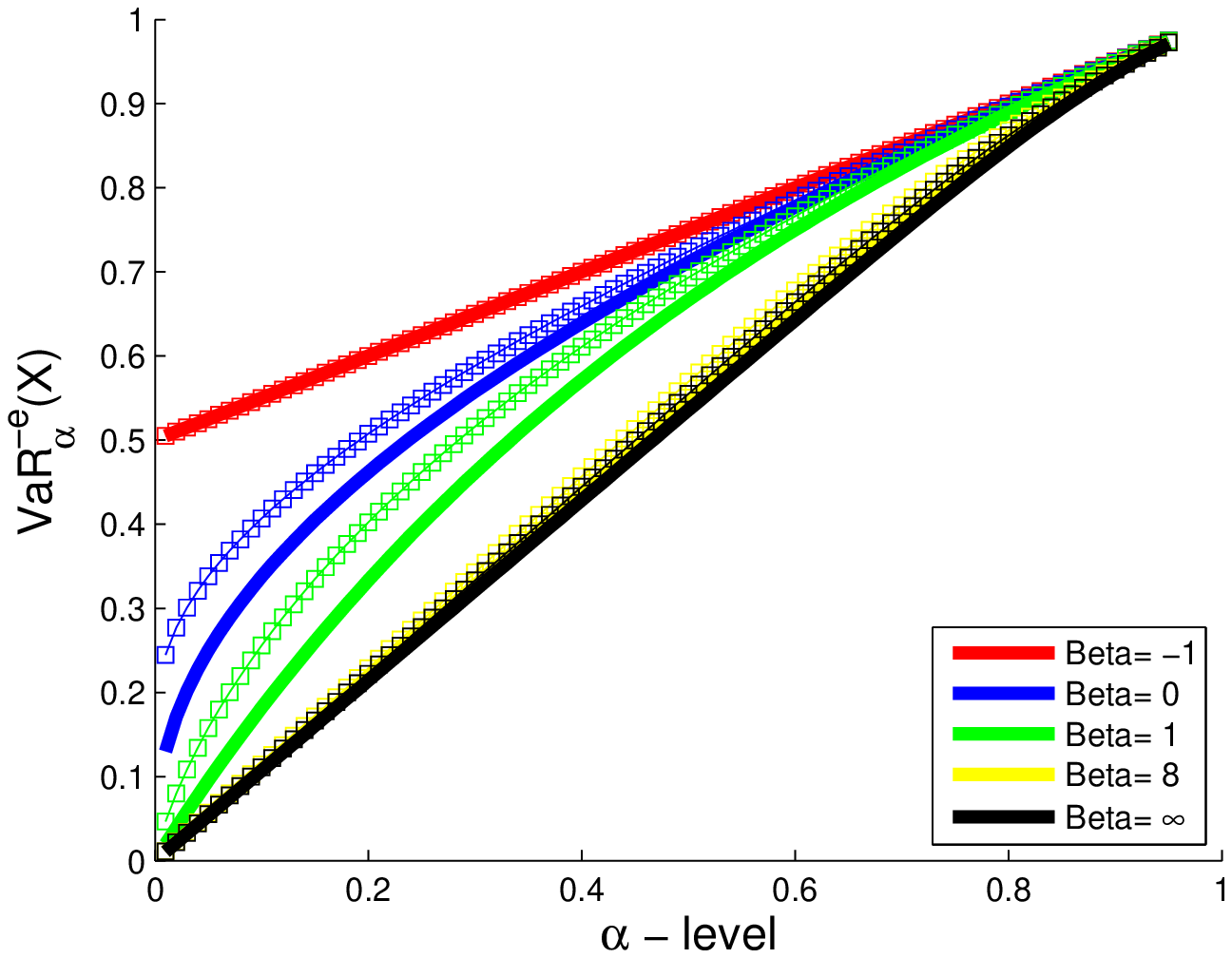}\hspace{0.5cm}\includegraphics[height=45mm,width=60mm]{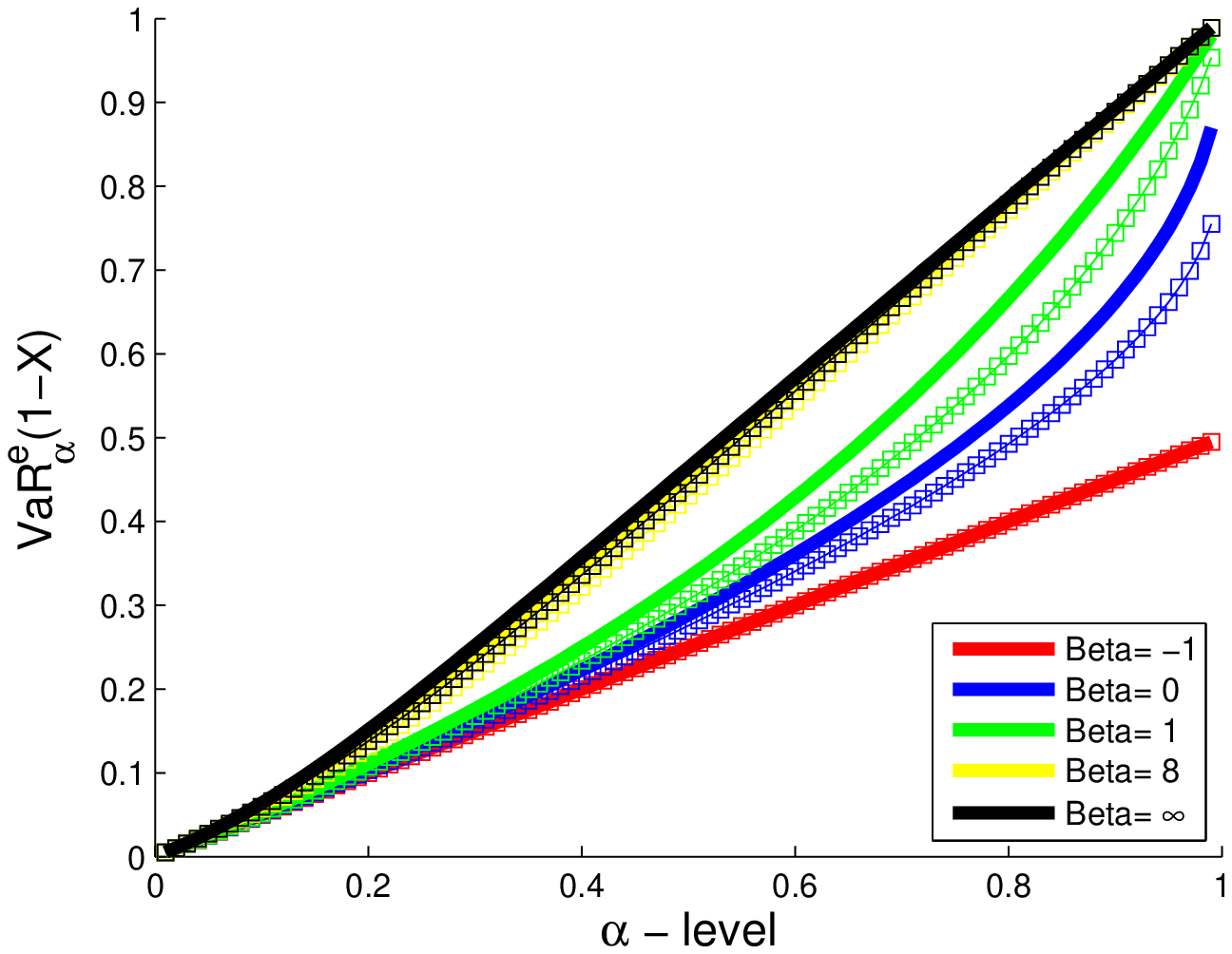}\ \ \\
\centerline{a)\;\;Lower Case\hspace{4.2cm}b)\;\;Upper Case}
	\caption{Comparison for Clayton's family of copulas.}
	\label{fig:copArqMRVaR}
\end{center}
\end{figure}

Table \ref{table:copVaR} contains the explicit expressions of $\underline{VaR}_{\alpha}(\mathbf{X})$ and $\overline{VaR}_{\alpha}(\mathbf{1-X})$ in dimension $2$, and the generalized expressions for our proposal in terms of $\alpha$, $\beta$ in any dimension. Figure \ref{fig:copArqMRVaR} shows the graphical comparison for $n=2$; the left plot presents the results for $VaR_{\alpha}^{-\mathbf{e}}(\mathbf{X})$ in solid line and $\underline{VaR}_{\alpha}(\mathbf{X})$ in dashed line, while the right plot presents the results for $VaR_{\alpha}^{\mathbf{e}}(\mathbf{1-X})$ in solid line and $\overline{VaR}_{\alpha}(\mathbf{1-X})$ in dashed line.

\begin{center}
\begin{table}[htbp]
\centering
\begin{tabular}{||c|c||c||}\hline\hline
 & Directional $VaR_{\alpha}^{\mathbf{u}}(\cdot)$ & \cite{bernardino}'s \textit{VaR}\\ \hline
$\mathbf{X}$ & $\left(\frac{1+\alpha^{-\beta}}{n}\right)^{-\frac{1}{\beta}}$ & $\frac{\beta}{\beta-1}\frac{\alpha^{\beta}-\alpha}{\alpha^{\beta}-1}$\\ \hline
$\mathbf{1-X}$ & $1-\left(\frac{1+(1-\alpha)^{-\beta}}{n}\right)^{-\frac{1}{\beta}}$ & $1-\frac{\beta}{\beta-1}\frac{(1-\alpha)^{\beta}-(1-\alpha)}{(1-\alpha)^{\beta}-1}$\\ \hline\hline
\end{tabular}
\caption{Clayton's Copula Case}
\label{table:copVaR}
\end{table}
\end{center}
\vspace{-1cm}

The results in Figure \ref{fig:copArqMRVaR} also shows us that in the case of random vectors with Clayton copula class, $VaR_{\alpha}^{-\mathbf{e}}(\mathbf{X})$ increases with respect to the parameter $\alpha$ and decreases in the parameter $\beta$. On the other side,  $VaR_{\alpha}^{\mathbf{e}}(\mathbf{1-X})$ is an increasing function of the parameter $\alpha$, but also an increasing function of the dependence parameter $\beta$. These features for this class of copulas were commented on and proved by \cite{bernardino} and for our risk measure can be easily proved following the same scheme. In addition, we need to highlight that for each fixed  pair ($\alpha$, $\beta$), the following relationships hold,

\begin{equation}
\underline{VaR}_{\alpha}(\mathbf{X}) \leq VaR_{\alpha}^{-\mathbf{e}}(\mathbf{X}) \quad\text{ and }\quad VaR_{\alpha}^{\mathbf{e}}(\mathbf{1-X}) \leq \overline{VaR}_{\alpha}(\mathbf{1-X}),
\end{equation}

where the inequalities are componentwise. Hence, we can say that our measurement is more conservative in the upper case and we are more optimistic in the lower case. This can be taken into consideration by the manager according to her/his preferences.

\vspace{-1cm}

\section{Robustness}\label{sec:rob}

The previous section presents the analytic results for random vectors with $[0,1]$-uniform marginals distributions. However,  in practical situations, it is necessary to obtain $VaR_{\alpha}^{\mathbf{u}}(\mathbf{X})$ for any random vector $\mathbf{X}$. In this case, we use the following computational approach summarized in the following pseudo-algorithm:

\begin{quotation}
	Input: $\mathbf{u}$, $\alpha$, $h$ and the multivariate sample $\mathbf{X}_{m}$.
	
	$\text{ for } i=1 \text{ to } m$\vspace{1pt}
	
	$\quad P_{i}=\mathbb{P}_{\mathbf{X}_{m}}\left[\mathfrak{C}
	_{\mathbf{x}_{i}}^{\mathbf{u}}\right]$,\vspace{1pt}
	
	$\quad \text{If } |P_{i}-\alpha|\leq h$\vspace{1pt}
	
	$\qquad \mathbf{x}_{i}\in \hat{\mathcal{Q}}_{\mathbf{X}_{m}}^{h}(\alpha,\mathbf{u})$,\vspace{1pt}
	
	$\quad$ end\vspace{1pt}
	
	$\quad \text{ for } \mathbf{x}_{j}\in \hat{\mathcal{Q}}_{\mathbf{X}_{m}}^{h}(\alpha,\mathbf{u})$\vspace{1pt}
	
	$\qquad d_{j} = dist(\mathbf{x}_{j},\{\boldsymbol{\mu}_{\mathbf{X}_{m}}+\lambda \mathbf{u}\})$,\vspace{1pt}
	
	$\quad$ end\vspace{1pt}
	
	end\vspace{1pt}
	
	$VaR_{\alpha}^{\mathbf{u}}(\mathbf{X}_{m})=\{\mathbf{x}_{k} | d_{k}=\min\{d_{j}\}\},$

\end{quotation}

where $\mathbf{X}_{m}:=\{\mathbf{x_{1}},\cdots,\mathbf{x_{m}}\}$ is the sample of the random vector $\mathbf{X}$, $\boldsymbol{\mu}_{\mathbf{X}_{m}}$ the sample mean, $\hat{\mathcal{Q}}_{\mathbf{X}_{m}}^{h}(\alpha,\mathbf{u}):=\left\lbrace \mathbf{x_{j}} : |\mathbb{P}_{\mathbf{X}_{m}}\left[\mathfrak{C}_{\mathbf{x_{j}}}^{\mathbf{u}}\right]-\alpha|\leq h\right\rbrace$ the sample quantile curve with a slack $h$ and $\mathbb{P}_{\mathbf{X}_{m}}[\cdot]$ is the empirical probability distribution of $\mathbf{X}_{m}$. Using this procedure, we are able to deal with high dimension random vectors. We are aware that this procedures can be improved using more sophisticated tools of the non-parametric statistics, but they are material for another project.

On the other hand,  it is well known that in risk theory, it is desirable that a measure be robust, (see \cite{artzner, burgert, cardin, rachev}). But in general, most of the measures are sensitive to atypical observations. In this section, we present a simulation study in order to describe the sensitivity of our proposal, using the  \cite{bernardino}'s measure as a benchmark.

The contamination model that we will use in the simulations is the following:

\begin{equation}\label{mixSim}
	\mathbf{X}^{\omega}\stackrel{\text{d}}{=}
	\begin{cases}
	\mathbf{X}_{1} \quad \text{ with probability $p=1-\omega$},\\
	\mathbf{X}_{2} \quad \text{ with probability $p=\omega$},
	\end{cases}
\end{equation}

where $\mathbf{X}_{1}\stackrel{\text{d}}{=}N_{1}(\boldsymbol{\mu}_{1},\Sigma_{1})$, $\mathbf{X}_{2}\stackrel{\text{d}}{=}N_{2}(\boldsymbol{\mu}_{1}+\Delta_{\boldsymbol{\mu}},\Sigma_{1}+\Delta_{\boldsymbol{\Sigma}})$ and $0 \leq \omega \leq 1$. The parameters of  $\mathbf{X}_{1}$ are,
\[\boldsymbol{\mu}_{1}=[50,50]', \qquad \Sigma_{1}=\begin{pmatrix}
0.5 & 0.3\\
0.3 &0.5
\end{pmatrix}.\]

 $\mathbf{X}_{1}$ remains fixed in the analysis, but the parameters of the normal distribution of $\mathbf{X}_{2}$ are changed to different steps to  generate outliers. As a measure to quantify  the effect of the outliers, we define,

\[PV^{\omega}=\frac{||Measure(\mathbf{X}^{\omega})-Measure(\mathbf{X}^{0})||_{2}}{||Measure(\mathbf{X}^{0})||},\]

where $Measure(\mathbf{X}^{0})$ is the risk measure evaluated in $\mathbf{X}^{0}$, with $\omega=0$ and $Measure(\mathbf{X}^{\omega})$ is a risk measure evaluated with the sample with a level of contamination $\omega\%$. 

\begin{center}
\begin{table}[htbp]
\centering
\begin{tabular}{||c|c||}\hline\hline
 Scenarios & Parameters of $\mathbf{X}_{2}$ distribution\\ \hline
Variance Analysis & $\boldsymbol{\mu}_{1}$,\quad $\Sigma_{1}+\begin{bmatrix}
4.5 & 0\\
0 & 6.5
\end{bmatrix}$\\ \hline
Covariance Matrix Analysis & $\boldsymbol{\mu}_{1}$,\quad $\Sigma_{1}+\begin{bmatrix}
4.5 & 0.2\\
0.3 & 6.5
\end{bmatrix}$\\ \hline
Mean Analysis & $\boldsymbol{\mu}_{1}+\Delta_{\boldsymbol{\mu}}$,\quad $\Sigma_{1}$\\ \hline
Join Analysis & $\boldsymbol{\mu}_{1}+\Delta_{\boldsymbol{\mu}}$, $\Sigma_{1}+\begin{bmatrix}
4.5 & 0.2\\
0.3 & 6.5
\end{bmatrix}$\\ \hline\hline
\end{tabular}
\caption{Simulation Stages and Parameters}
\label{table:robust}
\end{table}
\end{center}
\vspace{-1cm}

We have considered the scenarios for $\mathbf{X}_2$,  described in Table \ref{table:robust}. The procedure is the following: firstly, we have generated a non-contaminated sample $\mathbf{X}_{\omega}$, $\omega=0$ with 5000 observations and we calculate both $VaR_{0.1}^{\mathbf{e}}(\mathbf{X})$ and $\overline{VaR}_{0.1}(\mathbf{X})$.

Secondly, we have used the contamination model (\ref{mixSim}) taking values for $\omega$ from $1\%$ to $10\%$. Then, we generated for each $\omega$, $5000$ samples of $\mathbf{X}^{1}$ with an expected value of outliers $\omega\%$. We have  evaluated the risk measure as well as the percentage of variation for each level of contamination, performing this procedure $100$ times and we have reported the average of $PV^{\omega}$ in the following plots.
\begin{figure}[htbp]
\begin{center}
\includegraphics[height=45mm,width=45mm]{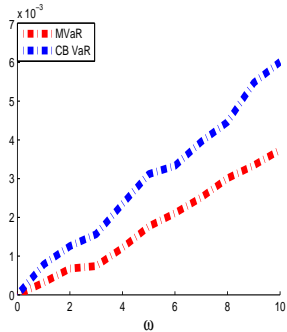}
    \caption{Percentage of variation of the measures varying the variances}\label{fig:robustVar}
\end{center}
\end{figure}
The first scenario suggests outliers given by changes on the variance of the marginals, which are difficult to detect in practice. We can see in Figure \ref{fig:robustVar} that the behavior of $VaR_{0.1}^{\mathbf{e}}(\mathbf{X})$ is better than that corresponding to \textit{upper-VaR} in \cite{bernardino} for any level of contamination. "Better", in this context, means that $PV^{\omega}$ is smaller.
\begin{figure}[htbp]
\begin{center}
\includegraphics[height=45mm,width=45mm]{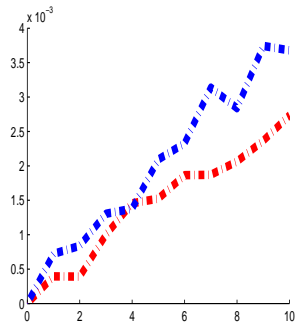}
    \caption{Percentage of variation of the measures varying the covariance matrix}\label{fig:robustCov}
\end{center}
\end{figure}
The second scenario considers changes in all the components of the covariance matrix. The results are reported in Figure \ref{fig:robustCov} that shows again the better behaviour of  $VaR_{0.1}^{\mathbf{e}}(\mathbf{X})$ with respect to robustness. The last scenarios consist of changes in the mean. Firstly, we affected the first component of the mean and then we affected the second one and finally both of them simultaneously.

\begin{figure}[htbp]
\includegraphics[height=30mm,width=30mm]{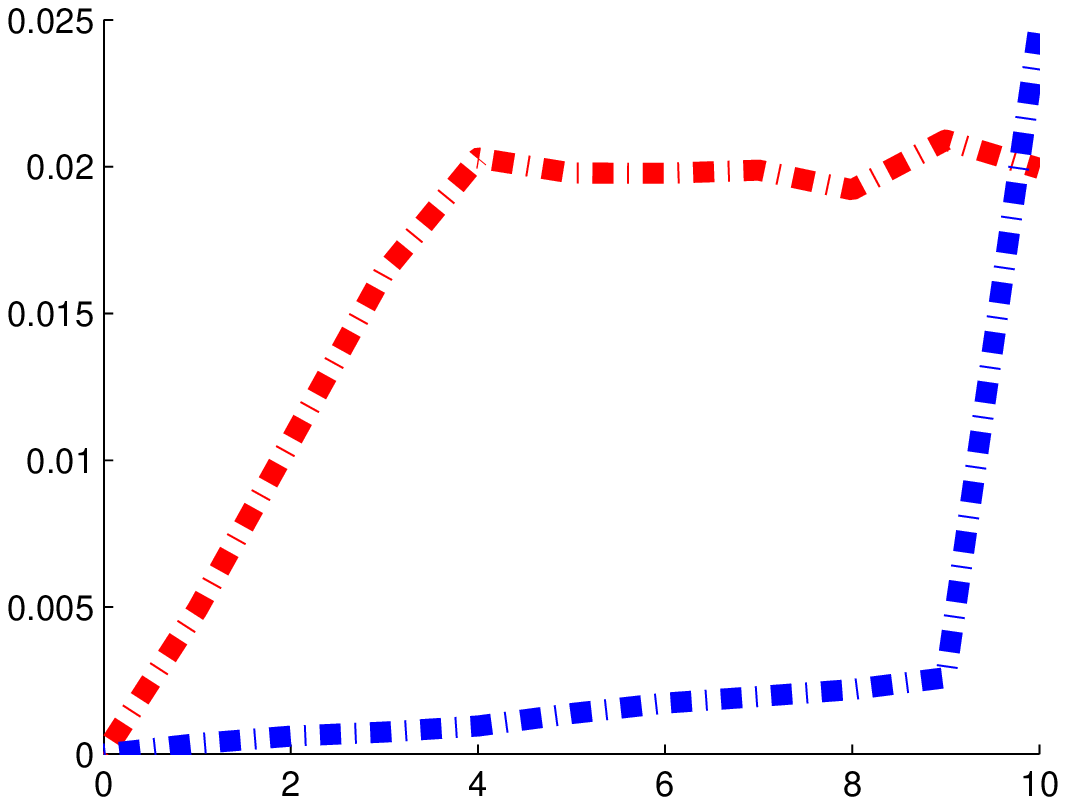}\hspace{1.5cm}\includegraphics[height=30mm,width=30mm]{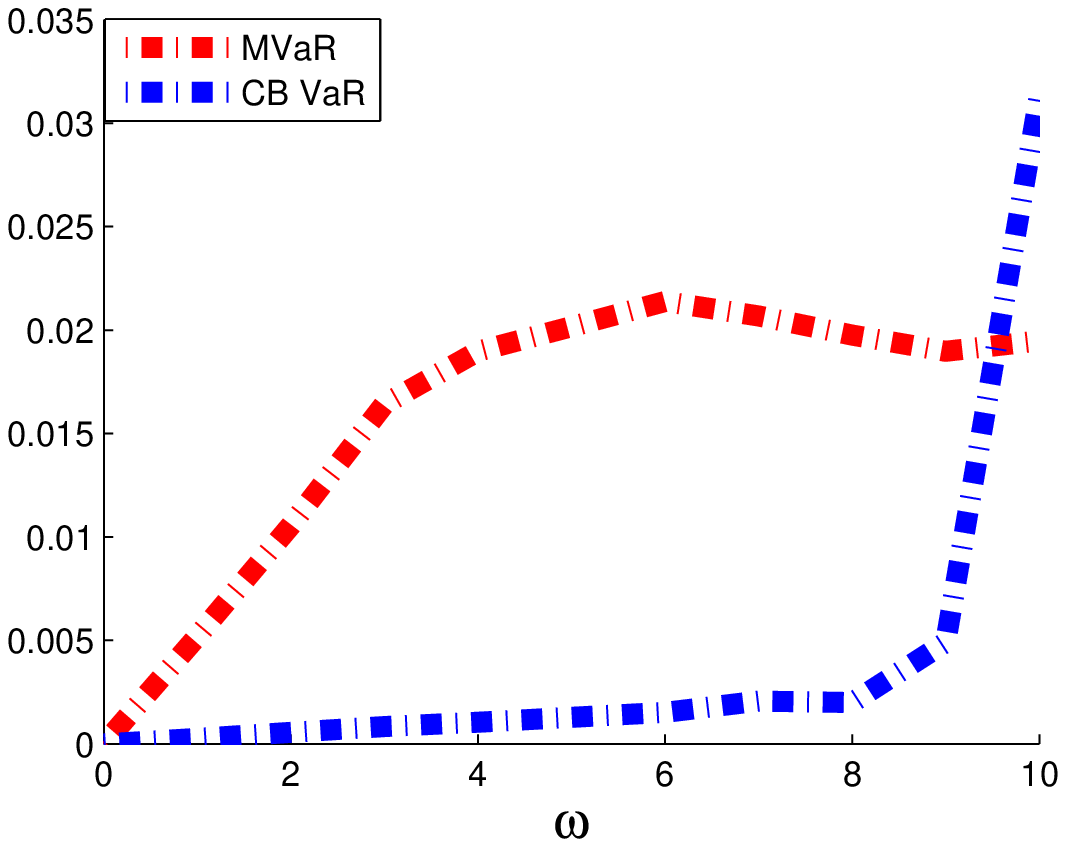}\hspace{1.5cm}\includegraphics[height=30mm,width=30mm]{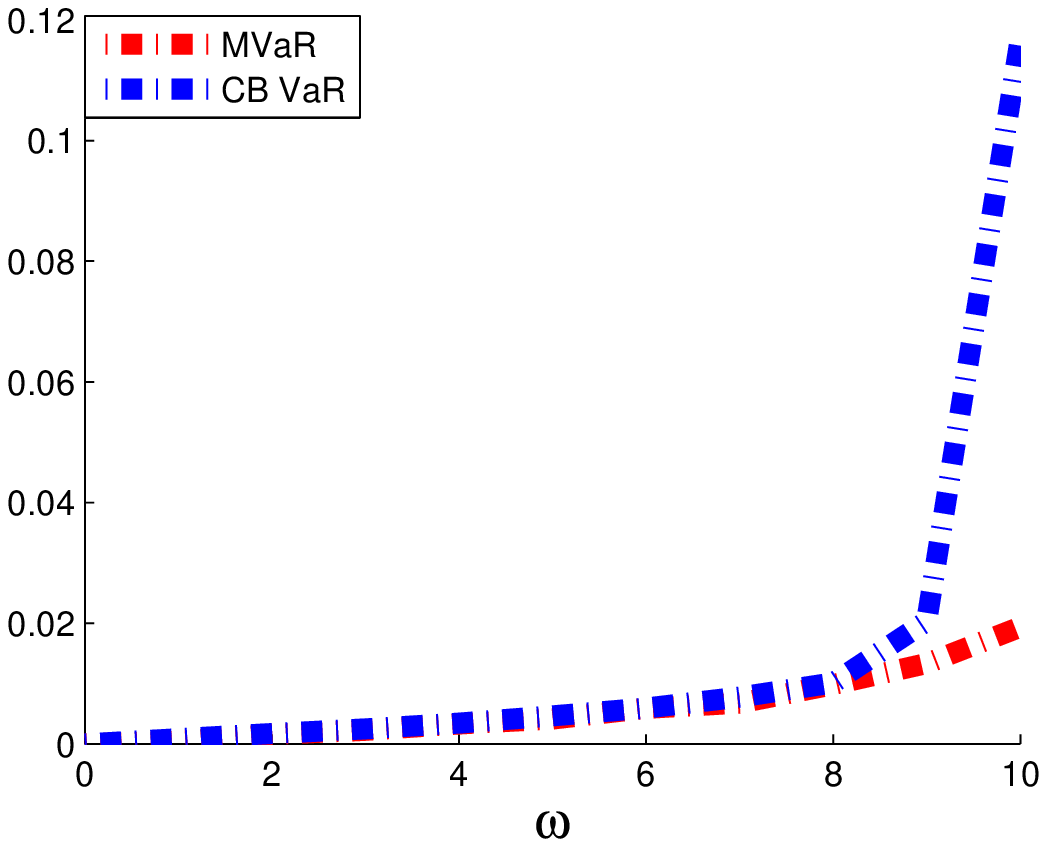}\ \ \\
\centerline{a)\;\;$\Delta_{\mu}=[0,25]'$\hspace{2.1cm}b)\;\;$\Delta_{\mu}=[25,0]'$\hspace{2cm}c)\;\;$\Delta_{\mu}=[25,25]'$}
    \caption{Percentage of variation of the measures varying all the parameters}\label{fig:robustAll}
\end{figure}

Figure \ref{fig:robustAll} summarizes  the results. As we can see,  $VaR_{0.1}^{\mathbf{e}}(\mathbf{X})$ shows robustness under the  presence of outliers of high dimension, but an extra-sensitivity under outliers in a unique component. The use of the mean of the random loss as the central point in the definition of our $VaR$ could be the cause of this lack of robustness.

\section{Conclusions}\label{sec:concl}

In this paper we have defined a multivariate extension of the classical risk measure \textit{VaR} based on a directional multivariate quantile recently introduced in the literature.
Specifically, we have proposed the \textit{directional multivariate Value at Risk} ($VaR_{\alpha}^{\mathbf{u}}(\mathbf{X})$) as a tool to analyze a portfolio of $n$ heterogeneous and dependent risks
considering external information or manager preferences.

We have analyzed the analytic properties of $VaR_{\alpha}^{\mathbf{u}}(\mathbf{X})$ in the same way as the \cite{artzner}'s axiomatic. We have provided some invariance properties as well as consistency and tail subadditivity property, which are desirable in a risk  measure.
We have shown relations between the components of the output of
$VaR_{\alpha}^{\mathbf{u}}(\mathbf{X})$ with respect to the corresponding univariate \textit{VaR} over the marginals.
A link between the univariate \textit{VaR} over the linear transformation using the portfolio weights vector $\mathbf{w}$, and the value of this transformation over $VaR_{\alpha}^{-\frac{\mathbf{w}}{||\mathbf{w}||}}(\mathbf{X})$ is given. We have also presented closed expressions for  $VaR_{\alpha}^{\mathbf{u}}(\mathbf{X})$ in terms of some families of copulas, considering particular dimensions or particular directions.

Finally we have presented a simulation study of robustness comparing the behavior of $VaR_{\alpha}^{\mathbf{u}}(\mathbf{X})$ with respect to  the  risk measure proposed in \cite{bernardino}. The simulations  show advantages of our proposal in relation to the presence of outliers. We have also detected in this study an open question to be taken into consideration in future work. The idea is to consider another central point instead of the mean as the center of the reference system, in order to improve the robustness of the risk measure, but, at the same time, keeping the good properties that have been proved. One option is to use a multivariate depth measure to choose the central point.

\section*{Acknowledgements}

This research was partially supported by a Spanish Ministry of Economyand Competition grant ECO2012-38442.

\section*{Appendix}

\begin{Proof}[\textbf{Proof of Lemma} \ref{prop:oddOrt}]
Since,

\begin{equation}\label{relMatrix}
	R_{\mathbf{u}}\mathbf{u}=\mathbf{e} \text{ and } R_{-\mathbf{u}}(-\mathbf{u})=\mathbf{e},
\end{equation}

we have that $R_{\mathbf{u}}=-R_{-\mathbf{u}}$. Then using (\ref{relMatrix}), we have that,
\[\begin{aligned}
		\mathfrak{C}_{\mathbf{x}}^{\mathbf{u}}&=\{\mathbf{z}\in\mathbb{R}^{n}:R_{\mathbf{u}}(\mathbf{z}-\mathbf{x})\geq 0\}\\
		&=\{\mathbf{z}\in\mathbb{R}^{n}:R_{-\mathbf{u}}(-\mathbf{z}-(-\mathbf{x}))\geq 0\}\\
		&=-\mathfrak{C}_{-\mathbf{x}}^{-\mathbf{u}}.
	\end{aligned}\]
\end{Proof}

\begin{Proof}[\textbf{Proof of Property} \ref{prop:oddMeasure}]
Due to Lemma \ref{prop:oddOrt}, it is easy to prove that
\begin{equation}\label{oddQuantile}
	\mathcal{Q}_{-\mathbf{X}}(\alpha,\mathbf{u})=-\mathcal{Q}_{\mathbf{X}}(\alpha,-\mathbf{u}),
\end{equation}

and hence,

\[\begin{aligned}
\mathcal{Q}_{-\mathbf{X}}(\alpha,\mathbf{u})\bigcap\{\lambda\mathbf{u}+\mathbb{E}[-\mathbf{X}]\}&\equiv \left(-\mathcal{Q}_{\mathbf{X}}(\alpha,-\mathbf{u})\right)\bigcap\left(-\{\lambda(-\mathbf{u})+\mathbb{E}[\mathbf{X}]\}\right)\\
&\equiv -\left(\mathcal{Q}_{\mathbf{X}}(\alpha,-\mathbf{u})\bigcap\{\lambda(-\mathbf{u})+\mathbb{E}[\mathbf{X}]\}\right).\\
\end{aligned}\]

Then,

\[VaR_{\alpha}^{\mathbf{u}}(-\mathbf{X})=-VaR_{\alpha}^{-\mathbf{u}}(\mathbf{X}).\]
\end{Proof}

\begin{Proof}[\textbf{Proof of Property} \ref{prop:homTrans}]
This property is derived  using Lemma \ref{prop:invOrt}.
\end{Proof}

\begin{Proof}[\textbf{Proof of Property} \ref{prop:consistencia}]
Since $\mathbf{X}\leq_{\mathcal{E}_{u}}\mathbf{Y}\Leftrightarrow R_{\mathbf{u}}\mathbf{X}\leq_{uo}R_{\mathbf{u}}\mathbf{Y}$, we get:
\small
\[L_{\mathbf{X}}(\alpha,\mathbf{u}):=\{\mathbf{z}\in\mathbb{R}^{n}:\mathbb{P}_{\mathbf{X}}(\mathfrak{C}_{\mathbf{z}}^{\mathbf{u}})\leq\alpha\}\supseteq \{\mathbf{z}\in\mathbb{R}^{n}:\mathbb{P}_{\mathbf{Y}}(\mathfrak{C}_{\mathbf{z}}^{\mathbf{u}})\leq\alpha\}:=L_{\mathbf{Y}}(\alpha,\mathbf{u})\]
\normalsize

Besides, $VaR_{\alpha}^{\mathbf{u}}(\mathbf{X})=\partial L_{\mathbf{X}}(\alpha,\mathbf{u})\bigcap \{\lambda\mathbf{u}+\mathbb{E}[\mathbf{X}]\}$ and $VaR_{\alpha}^{\mathbf{u}}(\mathbf{Y})=\partial L_{\mathbf{Y}}(\alpha,\mathbf{u})\bigcap \{\lambda\mathbf{u}+\mathbb{E}[\mathbf{Y}]\}$. Therefore, using the partial order defined in (\ref{porder}) there are three possibilities for $\mathbf{s},\mathbf{t}\in \mathbb{R}^{n}$:

\begin{enumerate}[(i)]
\item $\mathbf{s}\succ_{\mathbf{u}} \mathbf{t}$,
\item $\mathbf{s} \not\preceq_{\mathbf{u}} \mathbf{t}$ and $\mathbf{t} \not\preceq_{\mathbf{u}} \mathbf{s}$,
\item $\mathbf{s}\preceq_{\mathbf{u}} \mathbf{t}$.
\end{enumerate}

We can prove that the two first options are not possible for the points $VaR_{\alpha}^{\mathbf{u}}(\mathbf{X})$ and $VaR_{\alpha}^{\mathbf{u}}(\mathbf{Y})$. Suppose that
\[VaR_{\alpha}^{\mathbf{u}}(\mathbf{X}) \succ_{\mathbf{u}} VaR_{\alpha}^{\mathbf{u}}(\mathbf{Y}), \]
which implies that,
\[\mathfrak{C}_{\mathbf{z}_{\mathbf{X}}}^{\mathbf{u}}\subset \mathfrak{C}_{\mathbf{z}_{\mathbf{Y}}}^{\mathbf{u}}.\]
Hence,
\[\mathbb{P}_{\mathbf{Y}}(\mathfrak{C}_{\mathbf{z}_{\mathbf{Y}}}^{\mathbf{u}})\geq\mathbb{P}_{\mathbf{X}}(\mathfrak{C}_{\mathbf{z}_{\mathbf{Y}}}^{\mathbf{u}})> \mathbb{P}_{\mathbf{X}}(\mathfrak{C}_{\mathbf{z}_{\mathbf{X}}}^{\mathbf{u}})=\alpha.\]
In which case we arrive at a contradiction,if we assume the \textit{regularity conditions}. Moreover,
the hypothesis $\mathbb{E}[\mathbf{Y}]=c\mathbf{u}+\mathbb{E}[\mathbf{X}]$, for all $c>0$ and the conclusion  $\mathbb{E}[\mathbf{X}] \preceq_{\mathbf{u}} \mathbb{E}[\mathbf{Y}]$ derived in \cite{laniado-lillo-romo} (Property 3.4.), permits us to reject the second possibility of ordering between the two points. Thus,
the only option possible is,
\[VaR_{\alpha}^{\mathbf{u}}(\mathbf{X})\preceq_{u} VaR_{\alpha}^{\mathbf{u}}(\mathbf{Y})\]
\end{Proof}

\begin{Proof}[\textbf{Proof of Property} \ref{prop:rot}]
First, note  that:
\[\{\lambda(Q\mathbf{u})+\mathbb{E}[Q\mathbf{X}]\}=Q\{\lambda\mathbf{u}+\mathbb{E}[\mathbf{X}]\}.\]

Besides, we have the following relationship:

\begin{align*}
	\mathfrak{C}_{Q\mathbf{x}}^{Q\mathbf{u}}=\{z\in\mathbb{R}^{n}:R_{Q\mathbf{u}}(\mathbf{z}-Q\mathbf{x})\geq 0\}\quad \text{ and }\quad R_{Q\mathbf{u}}(Q\mathbf{u})=(R_{Q\mathbf{u}}Q)\mathbf{u}=\mathbf{e}.
\end{align*}

Then $R_{\mathbf{u}}=R_{Q\mathbf{u}}Q$, which implies that $\mathfrak{C}_{Q\mathbf{x}}^{Q\mathbf{u}}=Q\mathfrak{C}_{\mathbf{x}}^{\mathbf{u}}$,  and  $\mathbb{P}_{Q\mathbf{X}}(\mathfrak{C}_{Q\mathbf{x}}^{Q\mathbf{u}})=\mathbb{P}_\mathbf{X}(\mathfrak{C}_{\mathbf{x}}^{\mathbf{u}})$. Then, we get

\begin{equation}\label{rotQuantile}
	\mathcal{Q}_{Q\mathbf{X}}(\alpha,Q\mathbf{u})=Q\mathcal{Q}_{\mathbf{X}}(\alpha,\mathbf{u}),
\end{equation}
which proves the result.
\end{Proof}

\begin{Proof}[\textbf{Proof of Property} \ref{prop:nonExc}]
Property \ref{prop:rot} implies that,
\[R_{\mathbf{u}}VaR_{\alpha}^{\mathbf{u}}(\mathbf{X})=VaR_{\alpha}^{\mathbf{e}}(R_{\mathbf{u}}\mathbf{X}),\]
where $\mathbf{e}=\frac{\sqrt{n}}{n}[1,...,1]'$. Then,
\[R_{\mathbf{u}}VaR_{\alpha}^{\mathbf{u}}(\mathbf{X})\leq \sup_{\omega\in\Omega}\{R_{\mathbf{u}} \mathbf{X}(\omega)\},\]
and the proof is complete.
\end{Proof}

\begin{Proof}[\textbf{Proof of Property} \ref{prop:sub}]
It is easy to see that the equality in the mean implies that the vectors $\mathbb{E}[R_{\mathbf{u}}\mathbf{X}]$, $\mathbb{E}[R_{\mathbf{u}}\mathbf{Y}]$ and $\mathbb{E}[R_{\mathbf{u}}(\mathbf{X}+\mathbf{Y})]$ lie on the same line with direction vector $\mathbf{e}$. Then, we can write:

\begin{equation}\label{factorOrt}
\begin{aligned}
	\mathfrak{C}_{VaR_{\alpha}^{\mathbf{e}}(R_{\mathbf{u}}\mathbf{X})}^{\mathbf{e}} & = n[VaR_{\alpha}^{\mathbf{e}}(R_{\mathbf{u}}\mathbf{X})]_{1}\mathfrak{C}_{\mathbf{w}}^{\mathbf{e}}\\
	& = n[VaR_{\alpha}^{\mathbf{e}}(R_{\mathbf{u}}\mathbf{X})]_{1}[\mathbf{w},\infty)^{n},
\end{aligned}
\end{equation}

where $\mathbf{w}$ is the vector whose components are $\frac{1}{n}$ and $[\cdot]_{1}$ denotes the first component of the vector.

For $\alpha>0$ small, $\frac{1}{\alpha}\rightarrow \infty$, and then,

\[\frac{1}{\alpha}\mathbb{P}\left[(R_{\mathbf{u}}\mathbf{X},R_{\mathbf{u}}\mathbf{Y})\in \phi\left(\frac{1}{\alpha}\right)B\right]\rightarrow \mu(B).\]

On the other hand, we have the fact that the Borel set $\left(\phi\left(\frac{1}{\alpha}\right)\right)^{-1}\mathfrak{C}_{VaR_{\alpha}^{\mathbf{e}}(R_{\mathbf{u}}\mathbf{X})}^{\mathbf{u}}\times (0,\infty)^{n}$ satisfies the following relation:

\[\frac{1}{\alpha}\mathbb{P}\left[(R_{\mathbf{u}}\mathbf{X},R_{\mathbf{u}}\mathbf{Y})\in \left(\phi\left(\frac{1}{\alpha}\right)\right)\left(\phi\left(\frac{1}{\alpha}\right)\right)^{-1}(\mathfrak{C}_{VaR_{\alpha}^{\mathbf{e}}(R_{\mathbf{u}}\mathbf{X})}^{\mathbf{e}}\times (0,\infty)^{n})\right]\rightarrow 1.\]

Or equivalently,
\[\mu\left\{\left(\phi\left(\frac{1}{\alpha}\right)\right)^{-1}(\mathfrak{C}_{VaR_{\alpha}^{\mathbf{e}}(R_{\mathbf{u}}\mathbf{X})}^{\mathbf{u}}\times (0,\infty)^{n})\right\}\sim 1.\]
Hence using (\ref{factorOrt}), we have:

\begin{equation}\label{subad1}
	[VaR_{\alpha}^{\mathbf{e}}(R_{\mathbf{u}}\mathbf{X})]_{1}\sim \left(\mu\left\{\left[\frac{1}{n},\infty\right)^{n}\times (0,\infty)^{n}\right\}\right)^{\frac{1}{\beta}}\phi\left(\frac{1}{\alpha}\right)n.
\end{equation}

In the same way,

\begin{equation}\label{subad2}
	[VaR_{\alpha}^{\mathbf{e}}(R_{\mathbf{u}}\mathbf{Y})]_{1}\sim \left(\mu\left\{(0,\infty)^{n}\times \left[\frac{1}{n},\infty\right)^{n}\right\}\right)^{\frac{1}{\beta}}\phi\left(\frac{1}{\alpha}\right)n.
\end{equation}

Now,  in the case of the random variable $R_{\mathbf{u}}(\mathbf{X}+\mathbf{Y})$,
we have;
\begin{equation}\label{borelSum}
\begin{aligned}
\mathfrak{C}_{VaR_{\alpha}^{\mathbf{e}}\left(R_{\mathbf{u}}\left(\mathbf{X}+\mathbf{Y}\right)\right)}^{\mathbf{e}}&=
\{(\mathbf{x}, \mathbf{y})\in (0,\infty)^{2n}: (\mathbf{x}+\mathbf{y})> VaR_{\alpha}^{\mathbf{e}}(R_{\mathbf{u}}(\mathbf{X}+\mathbf{Y}))\}\\
&=n[VaR_{\alpha}^{\mathbf{e}}(R_{\mathbf{u}}(\mathbf{X}+\mathbf{Y}))]_{1}\cdot \{(\mathbf{x}, \mathbf{y})\in (0,\infty)^{2n}: (\mathbf{x}+\mathbf{y})> \mathbf{w}\}\\
&=n[VaR_{\alpha}^{\mathbf{e}}(R_{\mathbf{u}}(\mathbf{X}+\mathbf{Y}))]_{1}\cdot \mathfrak{C}_{\mathbf{w}}^{\mathbf{e}}
\end{aligned}
\end{equation}
where the inequalities in the expression are componentwise. As a consequence we get,


\[\mu\left\{\left(\phi\left(\frac{1}{\alpha}\right)\right)^{-1}\left\{(\mathbf{x}, \mathbf{y})\in (0,\infty)^{2n}: (\mathbf{x}+\mathbf{y})> VaR_{\alpha}^{\mathbf{e}}(R_{\mathbf{u}}(\mathbf{X}+\mathbf{Y}))\right\}\right\}\sim 1.\]

Then using the last equality in (\ref{borelSum}), we finally get,
\begin{equation}\label{subad3}
[VaR_{\alpha}^{\mathbf{e}}(R_{\mathbf{u}}(\mathbf{X}+\mathbf{Y}))]_{1}\sim \left(\mu\left\{\{(\mathbf{x}, \mathbf{y})\in (0,\infty)^{2n}: (\mathbf{x}+\mathbf{y})> \mathbf{w}\}\right\}\right)^{\frac{1}{\beta}}\phi\left(\frac{1}{\alpha}\right)n.
\end{equation}

Since in $\mathbb{R}^{n}$ all the norms are equivalent, i.e., for two norms $||\cdot||$ and $||\cdot||^{*}$, there are positive constants $c_{1},\ \ c_{2}$ such that $c_{1}||\cdot||\leq ||\cdot||^{*}\leq c_{2}||\cdot||$. Then, whatever norm is taken, we use the transformation [\cite{resnick1}, pg. 267.+], $\mathbf{x}\rightarrow (||\mathbf{x}||, ||\mathbf{x}||^{-1}\mathbf{x})$ and rewrite $\mu(\cdot)$ in terms of a new measure $\eta(\cdot)$ in $\mathcal{D}:=\{\mathbf{z}\in [0,\infty]^{2n}\backslash \{\mathbf{0}\}: ||\mathbf{z}||=1\}$ as $r^{-\beta}\eta(\cdot)$, due to the property of the measure in (\ref{regVarying}). The relationship satisfying both measures for a Borel set $A$ in $\mathcal{D}$, it is given by,

\begin{equation}\label{intTransform}
	\mu(A)=\int_{\mathcal{D}}\int_{0}^{\infty}\mathbf{1}(r(\mathbf{u},\mathbf{v})\in A)\beta r^{-(1+\beta)}dr\eta(d\mathbf{u},d\mathbf{v}).
\end{equation}

Then the measure of the Borel sets in (\ref{subad1}), (\ref{subad2}) and (\ref{subad3}) can be expressed using $||\cdot||_{1}$ as:

\begin{align}
	\mu\left(\left(\frac{1}{n},\infty\right)^{n}\times (0,\infty)^{n}\right)&=\int_{\mathcal{D}}\left(\sum_{i}u_{i}\right)^{\beta}\eta(d\mathbf{u},d\mathbf{v}),\\
	\mu\left((0,\infty)^{n}\times \left(\frac{1}{n},\infty\right)^{n}\right)&=\int_{\mathcal{D}}\left(\sum_{i}v_{i}\right)^{\beta}\eta(d\mathbf{u},d\mathbf{v}),\\
	\mu(\{(\mathbf{x}, \mathbf{y})\in (0,\infty)^{2n}: (\mathbf{x}+\mathbf{y})> \mathbf{w}\})&=\int_{\mathcal{D}}\left(\sum_{i}(u_{i}+v_{i})\right)^{\beta}\eta(d\mathbf{u},d\mathbf{v}),
\end{align}

Now using the Mikowski inequality we obtain:
\small
\begin{equation}\label{minkowsky}
	\left(\int_{\mathcal{D}}\left(\sum_{i}(u_{i}+v_{i})\right)^{\beta}\eta(d\mathbf{u},d\mathbf{v})\right)^{\frac{1}{\beta}}\leq \left(\int_{\mathcal{D}}\left(\sum_{i}u_{i}\right)^{\beta}\eta(d\mathbf{u},d\mathbf{v})\right)^{\frac{1}{\beta}}+\left(\int_{\mathcal{D}}\left(\sum_{i}v_{i}\right)^{\beta}\eta(d\mathbf{u},d\mathbf{v})\right)^{\frac{1}{\beta}}.
\end{equation}
\normalsize
Hence combining (\ref{subad1}), (\ref{subad2}), (\ref{subad3}) and (\ref{minkowsky}), we have the result

\[[VaR_{\alpha}^{\mathbf{e}}(R_{\mathbf{u}}(\mathbf{X}+\mathbf{Y}))]_{1}\leq [VaR_{\alpha}^{\mathbf{e}}(R_{\mathbf{u}}\mathbf{X})]_{1}+[VaR_{\alpha}^{\mathbf{e}}(R_{\mathbf{u}}\mathbf{Y})]_{1},\]

or equivalently, from Proposition \ref{prop:rot} and the partial order defined in (\ref{porder}), we have for $\mathbf{u}=\frac{\mathbf{m}}{||\mathbf{m}||}$ that:

\begin{equation}
	VaR_{\alpha}^{\mathbf{u}}(\mathbf{X}+\mathbf{Y})\preceq_{\mathbf{u}} VaR_{\alpha}^{\mathbf{u}}(\mathbf{X})+VaR_{\alpha}^{\mathbf{u}}(\mathbf{Y}).
\end{equation}
\end{Proof}

\begin{Proof}[\textbf{Proof of Proposition} \ref{prop:cotaHenry}]
By Definition \ref{qExt},  if $\mathbf{x}\in\mathcal{Q}_{\mathbf{X}}(\alpha,\mathbf{u})$, we have $\mathbb{P}[R_{\mathbf{u}}(\mathbf{X}-\mathbf{x}) \geq 0]=\alpha$.
Therefore,

\begin{equation}\label{eqAux}
\mathbb{P}[\mathbf{1}' R_{\mathbf{u}}(\mathbf{X}-\mathbf{x}) \geq 0] \geq \alpha \qquad\text{ where } \mathbf{1}=[1,\cdots,1]'.
\end{equation}

Since $R_{\mathbf{u}}\mathbf{u} = \mathbf{e}$, we obtain,

\begin{align*}
\mathbb{P}[\mathbf{1}' R_{\mathbf{u}}(\mathbf{X}-\mathbf{x}) \geq 0]  & = \mathbb{P}[\sqrt{n}(R_{\mathbf{u}}\mathbf{u})' R_{\mathbf{u}}(\mathbf{X}-\mathbf{x}) \geq 0] \\
& = \mathbb{P}[\sqrt{n}\left(-\frac{\mathbf{w}}{||\mathbf{w}||}\right)' (\mathbf{X}-\mathbf{x}) \geq 0]\\
& = \mathbb{P}[\mathbf{w}' \mathbf{X} \leq \mathbf{w}'\mathbf{x}] = \mathbb{P}[Z \leq \mathbf{w}'\mathbf{x}]
\end{align*}

Thus, (\ref{eqAux}) and (\ref{VaR}) implies $\mathbf{w}'\mathbf{x} \geq VaR_{\alpha}(Z)$.

\end{Proof}

\begin{Proof}[\textbf{Proof of Proposition} \ref{prop:relMarginal}]
The proof follows the same outline as that of [\cite{bernardino}, Proposition 2.4.]. Note that in direction $\mathbf{u}=\mathbf{e}$,
\[
\mathfrak{C}_{\mathbf{x}}^{\mathbf{e}}=\{\mathbf{z}\in\mathbb{R}^{n}:\mathbf{z}\geq\mathbf{x}\}.
\]

Then we can write,
\begin{align*}
L_{\alpha}&=\{\mathbf{x}\in\mathbb{R}^{n}:\mathbb{P}(\mathfrak{C}_{\mathbf{x}}^{\mathbf{e}})\leq \alpha\}\\
&=\{\mathbf{x}\in\mathbb{R}^{n}:\mathbb{P}(\mathbf{X}\geq \mathbf{x})\leq \alpha\}
\end{align*}

And we can assume  the convexity of the $[L_{\alpha}]^{c}$ by the quasi-concavity of the survival function $\bar{F}$, where $[\cdot]^{c}$ denotes the complementary set.
Now, as $\mathcal{Q}_{\mathbf{X}}(\alpha,\mathbf{e})= \partial L_{\alpha}\equiv \partial [L_{\alpha}]^{c}$,  $VaR_{\alpha}^{\mathbf{e}}(\mathbf{X})$ belongs to the set $\partial [L_{\alpha}]^{c}$. Moreover, from the definition of survival function we have that,
\small
\[\bar{F}(\infty,\cdots,x_{i},\cdots,\infty)\geq\bar{F}(\mathbf{x})=\bar{F}(x_{1},\cdots,x_{i},\cdots,x_{n})\quad\text{for all}\quad \mathbf{x}\in\mathbf{R}^{n}\text{ and } i=1,\cdots,n.\]

\normalsize
Then each component of a vector belonging to $\partial[L_{\alpha}(\mathbf{e})]^{c}$ is superiorly bounded by the univariate \textit{VaR} at level $p=1-\alpha$ of the corresponding marginal. As a consequence,
 each component of $VaR_{\alpha}^{\mathbf{e}}(\mathbf{X})$ is superiorly bounded by the univariate \textit{VaR} at level $p=1-\alpha$ of the corresponding marginal and hence, the first inequality holds. Now for the second inequality,

\[
\mathfrak{C}_{\mathbf{x}}^{-\mathbf{e}}=\{\mathbf{z}\in\mathbb{R}^{n}:\mathbf{z}\leq\mathbf{x}\}.
\]

Then, we have,

\begin{align*}
L_{1-\alpha}&=\{\mathbf{x}\in\mathbb{R}^{n}:\mathbb{P}(\mathfrak{C}_{\mathbf{x}}^{-\mathbf{e}})\leq 1-\alpha\}\\
&=\{\mathbf{x}\in\mathbb{R}^{n}:\mathbb{P}(\mathbf{X}\leq \mathbf{x})\leq 1-\alpha\}
\end{align*}

But, if $F$ is a quasi-concave function, we have that $[L_{1-\alpha}]^{c}$ is a convex set and $\mathcal{Q}_{\mathbf{X}}(1-\alpha,-\mathbf{e})=\partial L_{1-\alpha}\equiv \partial [L_{1-\alpha}]^{c}$. Therefore $VaR_{1-\alpha}^{\mathbf{e}}(\mathbf{X})$ belongs to the set $[L_{1-\alpha}]^{c}$. Additionally, from the definition of distribution function, it is easy to show that each component of an element in $[L_{1-\alpha}]^{c}$ is inferiorly bounded by the univariate \textit{VaR} at level $p=1-\alpha$ of the corresponding marginal; hence, we obtain the result.
\end{Proof}

\end{document}